\documentclass{article}
\usepackage[utf8]{inputenc}
\usepackage{geometry}
\usepackage{hyperref}
\usepackage{graphicx}
\usepackage{amsmath}
\usepackage{mathtools}
\usepackage{float}
\usepackage{amsfonts}
\usepackage{braket}
\usepackage{tikz}
\usepackage{cite}
\usepackage{appendix}
\newgeometry{tmargin=2cm, bmargin=3cm, lmargin=2cm, rmargin=2cm}
\newcommand\qed{$\Box$}
\renewcommand\d{\mathrm{d}}
\newcommand\e{\mathrm{e}}

\newcommand\p{\mathrm{p}}
\newcommand\cF{\mathcal{F}}

\newcommand\cD{\mathcal{D}}
\newcommand\rr{\mathbb{R}}

\newtheorem{theorem}{Theorem}[section]
\newtheorem{proposition}[theorem]{Proposition}

\newtheorem{lemma}[theorem]{Lemma}
\numberwithin{equation}{section}

\title{
  Momentum approach to  the  $1/r^2$ potential\\
  as a toy model of  the Wilsonian renormalization}
\author{Jan Derezi\'nski\\
	email \href{mailto:jan.derezinski@fuw.edu.pl}{jan.derezinski@fuw.edu.pl} \\ Oskar Grocholski\\
	email \href{mailto:o.grocholski@student.uw.edu.pl}{o.grocholski@student.uw.edu.pl}\\
Faculty of Physics, University of Warsaw,\\
 Warsaw, Poland}
\date{\today}

\begin{document}

\maketitle

\begin{abstract}
  The Bessel operator, that is, the
  Schr\"odinger operator on the halfline with a potential proportional to $1/x^2$, is analyzed in the momentum representation. Many features of this analysis are parallel to the approach  \`a la K.~Wilson  to Quantum Field Theory: one needs to impose a cutoff, add counterterms, study the renormalization group flow with its fixed points and limit cycles.
\end{abstract}


\section{Introduction}

Our paper is devoted to  one of the most curious families of operators in mathematical physics,
\begin{equation}
    L_\alpha \coloneqq -\frac{\d^2}{\d x^2} + \Big{(} \alpha - \frac{1}{4} \Big{)} \frac{1}{x^2},
    \label{eq-L-m-}
\end{equation}
where $\alpha\in\mathbb{R}$. 
 $L_\alpha$ are often called 
{\em  Bessel operators}.  
{ We will use the symbol $L_\alpha$ for the  expression
  \eqref{eq-L-m-}, without specifying its domain. Operators that are
  given by this expression and have a concrete domain will be called
  {\em realizations of $L_\alpha$}.}
  Let us briefly list their  most important properties.
  
Bessel operators are Hermitian (symmetric)
on  $C_\mathrm{c}^\infty(\rr_+)$\footnote{{ By $\rr_+$ we always denote the open positive half-line
$]0,\infty[$. Thus functions in
 $C^\infty_\mathrm{c}(\mathbb{R}_+)$ are supported away from $0$.}}. However, they are  essentially self-adjoint
    only  for $\alpha\geq1$.
For $\alpha<1$ they
possess a 1-parameter family of self-adjoint realizations.
In the terminology of Weyl, \cite{Weyl}, $0$ is {\em limit point} for $\alpha>1$ and 
{\em limit circle} for $\alpha\leq1$. 

For $\alpha\neq0$  their self-adjoint realizations
can be described by specifying  boundary conditions at zero as  appropriate mixtures of $x^{\frac12+m}$ and $x^{\frac12-m}$, where $m:=\sqrt{\alpha}$. For  $\alpha=0$, one needs to take mixtures  of  $x^\frac12$ and $x^{\frac12}\ln x$ and   there is a curious ``phase transition'': $m$ is real for $\alpha\geq0$ and imaginary for $\alpha<0$. Moreover,  all self-adjoint realizations of $L_\alpha$ are bounded from below for $\alpha\geq0$ and unbounded from below for $\alpha<0$.

 Many quantities related to    $L_\alpha$ can be computed
 explicitly. For instance,  the eigenfunction expansion
of  $L_\alpha$ can be given in terms of Bessel functions, see e.g. the classic book by Titchmarsh \cite{Tit}. This allows us to describe explicitly all
   self-adjoint realizations of $L_\alpha$,
as discussed by many authors, e.g.
  \cite{Meetz,AnanievaBudika, CoonHolstein, Pankrashkin, Kovarik, Griffiths, Case,GTV,Derezinski1, Derezinski2, Inoue}.

 It is useful to note that $d$-dimensional Schr\"odinger operators with the inverse square potential
 \begin{equation}\label{eq-L-m-d}
   L_{\alpha,d}:=-\Delta_d+\frac{g}{r^2}\end{equation}
 can be reduced to  $1$-dimensional ones (\ref{eq-L-m-}).
 In fact, if we restrict (\ref{eq-L-m-d}) to spherical harmonics of  degree $l$ and make a simple transformation, then the radial operator coincides with $L_\alpha$, where \begin{equation}{  \alpha=\Big(\frac{d}{2}-1+l\Big)^2.}\label{alpha}\end{equation}
In our paper, for  simplicity we always stick  to the dimension $1$, except for a short resume of the reduction of the $3$- to the $1$-dimensional case.

One of the most striking properties of Bessel operators is their homogeneity of degree $-2$. In other words, if $U_\tau$ denotes the scaling transformation (see (\ref{scaling})), then
\begin{equation}\label{scaling1}
 U_\tau L_\alpha U_\tau^{-1}= \e^{-2\tau}L_\alpha.
\end{equation}
This suggests  us to introduce the transformation
\begin{equation}\label{rg}
  R_\tau(B):= \e^{2\tau}U_\tau B U_\tau^{-1},\end{equation}
acting on, say, self-adjoint operators.  $\mathbb{R}\ni \tau\mapsto R_\tau$ is a representation of the group $\mathbb{R}$
and preserves the set of self-adjoint realizations of $L_\alpha$.
Some of these realizations are ``fixed points'' of $R_\tau$.
There is an interesting analogy between the action of
$R_\tau$ \eqref{rg} on 
self-adjoint extensions of
Bessel operators %$L_\alpha$
and  
the renormalization group acting on  models of
Quantum Field Theory
(QFT).

The most obvious approach to Bessel operators is to study them in
the position representation, as it is usually done in the
literature. Our paper is devoted to an analysis of Bessel operators in
the momentum representation. We believe that this is interesting in
itself---in fact, many things look very different on the momentum
side. Our main motivation, however, comes from the physics literature,
where  some authors try
to explain  the renormalization group approach to 
QFT by using Bessel operators  as a toy model.

In  QFT one can work both in the position  and in the
momentum space. The position representation is used in  the so-called
{\em Epstein-Glaser approach}. In practice, however, physicists prefer to
work in the momentum representation, which is usually more convenient
for computations.
%it is usually more convenient to work in the momentum  rather than   position representation. This is related to the fact that
%experimentalists  have  at their disposal a finite range of energies.
In order to compute various useful quantities one typically needs to
impose a  cutoff at momentum $\Lambda$ and to add appropriate
$\Lambda$-dependent counterterms. The desired quantities 
are obtained in the limit $\Lambda\to\infty$, provided that various
parameters are appropriately adjusted.

This procedure was greatly clarified by the Nobel prize winner
K.~Wilson, who stressed the role of scaling transformations in the
process of renormalization. He also pointed out that it is not so
important whether QFT has a well defined limit for $\Lambda\to\infty$:
what matters is a weak dependence of low energy quantities on  the high energy cutoff.

A number of authors \cite{Glazek, vanKolck, Long, Hammer, Braaten},
mostly with a high energy physics  background, noticed that Bessel
operators have a great pedagogical potential to illustrate the concept
of renormalization group and Wilson's ideas. In particular we would like to draw the
reader's attention to a recent paper \cite{Glazek},  which  is 
the main inspiration for our work. Reference \cite{Glazek}
looks at the operator  (\ref{eq-L-m-d}) in dimension 3 in the momentum representation. Its naive formulation is ill-defined due to diverging integrals at large momenta and
does not select a self-adjoint realization.
To cure these problems \cite{Glazek} applies the following steps,  parallel to the usual approach to QFT:
\begin{subequations}\label{subu}
\begin{align} &\text{Cut off the formal Hamiltonian with a {\em momentum 
  cutoff} $|p|<\Lambda$. }\\
 &\text{Add an appropriate {\em countertem} multiplied by a 
 {\em running   coupling constant}. }\\
 &\text{Determine the differential equation for the coupling constant. }\\
&\text{Take the limit $\Lambda\to\infty$. }
  \end{align}\end{subequations}
The exposition contained  in \cite{Glazek} stays all the time on the 
momentum side, and the reader may find  it difficult to connect it 
to the standard theory of self-adjoint realizations of Bessel operators, 
more transparent  in the position representation.

%In our paper we analyze rigorously Bessel operators in the
%momentum representation.
%We will see that  various concepts known from
%QFT, such as cutoffs, counterterms, running coupling constant and the renormalization group \`a la Wilson
%appear naturally in this analysis.

After describing the main source of motivation of our paper, let us
outline its content.
We start with  Sec. \ref{sec1} containing a brief resum\'{e} of the
theory of self-adjoint realizations of Bessel operators
% $L_\alpha$ on $L^2(\mathbb{R}_+)$
in the position representation. We  follow the terminology and conventions of \cite{Derezinski1, Derezinski2}.

In Sec.
\ref{sec2} we  recapitulate the content of \cite{Glazek}. On purpose,
we stick to the original terminology and line of reasoning. In
particular, we use the 3-dimensional setting, which can clearly be 
reduced to  1 dimension by the use of spherical coordinates.

Sec. \ref{sec3} is the main part of our work.
We start with a brief analysis of the momentum approach to general
Schr\"odinger operators. 
We 
are mostly interested in operators on  the
half-line $\mathbb{R}_+$,
and not on  the whole line $\mathbb{R}$.
This  is consistent with the
many-dimensional analysis in spherical coordinates,
where the radius is always positive. 
We encounter the following issue: the usual Fourier transformation is
adapted to the line, whereas on the half-line we have two natural
cousins of the Fourier transformation: the sine and the cosine
transformation {     (see \eqref{fd}, resp. \eqref{fn})}. The former diagonalizes the Dirichlet Laplacian
$-\Delta_\mathrm{D}$, and the latter the Neumann Laplacian
$-\Delta_\mathrm{N}$. Which approach we should choose? One can argue
that  from the
1-dimensional point of view both are equally natural.
Therefore, we discuss both approaches. Actually, from
the 3-dimensional point of view   the sine
transformation should be preferred, because the 3-dimensional Laplacian in
the 
s-wave sector reduces to the Dirichlet Laplacian on the half-line. 

Let us briefly describe the momentum  approach to Schr\"odinger
operators on $\rr_+$ for sufficiently nice potentials.
There are  two basic Schr\"odinger operators with the
potential $V$ on the halfline: $-\Delta_\mathrm{D}+V(x)$
and $-\Delta_\mathrm{N}+V(x)$.
By applying the sine transformation to the former and 
the cosine transformation to the latter we obtain 
the operators
\begin{align}
  \Tilde{H}_\mathrm{D}\psi(p)=p^2\psi(p)+\frac{1}{2\pi}\int_0^\infty(\Tilde{ V}(p-q)-\Tilde{ V}(p+q)\big)\psi(q)\mathrm{d} q,\label{momen1a}\\
    \Tilde{H}_\mathrm{N}\psi(p)=p^2\psi(p)+\frac{1}{2\pi}\int_0^\infty(\Tilde{ V}(p-q)+\Tilde{ V}(p+q)\big)\psi(q)\mathrm{d} q.\label{momen2a}
\end{align}
Here, $\Tilde{V}(p)$ is the Fourier transform of the 
even extension of $V$ from $\mathbb{R}_+$ to $\mathbb{R}$.

If the potential is singular at $0$, say, non-integrable, we have several problems:
\begin{enumerate}
\item We cannot use the standard Dirichlet/Neumann boundary conditions.
\item We need to interpret $V$ as an irregular distribution.
\item The Fourier transform of $V$ will grow at infinity, making the formulas (\ref{momen1a}) and (\ref{momen2a}) problematic.
\end{enumerate}
  
  All these problems are present for 
  Bessel operators, where $V$ is proportional to $\frac1{x^2}$.
  In particular,  $\frac1{x^2}$ does not define a regular distribution.
  There exists, however, a well-known one-parameter family of even distributions that outside of $0$ coincide with $\frac1{x^2}$. Their Fourier transforms are 
  \begin{equation}-\pi |p|+a,\label{oao}\end{equation}
  where $a$ is a constant.

  A priori it is not obvious which transformation is more appropriate
  for Bessel operators: the sine or the cosine transformation. Let us consider both, setting
\begin{align}
  \Tilde{  L}_{\alpha,\mathrm{D}}&:= \cF_\mathrm{D} L_\alpha\cF_\mathrm{D}
  ,\\
    \Tilde{  L}_{\alpha,\mathrm{N}}&:= \cF_\mathrm{N} L_\alpha \cF_\mathrm{N},
\end{align}
where $L_\alpha$ is treated as a formal expression, 
  $\cal{F}_{\mathrm{D}/\mathrm{N}}$ is the sine/cosine transformation.

Let us use $-\pi |p|$ as the Fourier transform of $\frac{1}{x^2}$.
(We set $a=0$ in \eqref{oao}. This constant will reappear in disguise of a   counter-term  anyway later.) 
(\ref{momen1a}) and (\ref{momen2a}) yield formal expressions
  \begin{align}
    \Tilde{  L}_{\alpha,\mathrm{D}}\psi(p)&
    =
    p^2\psi(p)+\Big(\alpha-\frac14\Big)\int_0^\infty\big(\theta(p-q)q
    +\theta(q-p)p\big)\psi(q)\mathrm{d}q
  ,\label{mm1a}\\
    \Tilde{  L}_{\alpha,\mathrm{N}}\psi(p)&
    =
    p^2\psi(p)-\Big(\alpha-\frac14\Big)\int_0^\infty\big(\theta(p-q)p
    +\theta(q-p)q\big)\psi(q)\mathrm{d}q,\label{mm2a}
  \end{align}
where  $\theta$ is the Heaviside function.
Both expressions are problematic for
large momenta. In what follows we give two constructions of
self-adjoint realizations of \eqref{mm1a} and \eqref{mm2a}.
The first construction will be called
``mathematicians' style'', and the second 
``physicists' style''.

The construction in ``mathematicians' style''  directly describes
the domains and actions
of the operators. It starts with a construction of the
{\em minimal Bessel operators}, denoted
 $  \Tilde{
  L}_{\alpha,\mathrm{D}}^{\min}$ and $   \Tilde{
  L}_{\alpha,\mathrm{N}}^{\min}$.
They are defined by  \eqref{mm1a} and \eqref{mm2a} on
  appropriate domains consisting of rapidly
decaying functions. In addition, for \eqref{mm1a} we need to assume
that $\int_0^\infty\psi(p)p\d p=0$ and for \eqref{mm2a}
$\int_0^\infty\psi(p)\d p=0$. 

The adjoints of
the minimal Bessel operators are called the {\em maximal Bessel operators},
and denoted
 $  \Tilde{
  L}_{\alpha,\mathrm{D}}^{\max}$ and $   \Tilde{
  L}_{\alpha,\mathrm{N}}^{\max}$.
If $\alpha\geq1$, they coincide with the  minimal Bessel operators,
and also with their unique self-adjoint realizations. This is not the
case for $\alpha<1$.
{   The construction of maximal operators in the momentum representation
is somewhat tricky. To obtain the maximal domains, $\cD(   \Tilde{
  L}_{\alpha,\mathrm{D}}^{\min})$ and $  \cD(  \Tilde{
  L}_{\alpha,\mathrm{N}}^{\min})$
have to be  extended by adjoining appropriate  2-dimensional
subspaces.
We cannot directly use
the expressions \eqref{mm1a} and  \eqref{mm2a}, since now
they usually contain divergent integrals.
 If $\alpha\neq\frac14$ (that is, when there is a
non-trivial potential), this additional subspace is spanned
by certain vectors that behave for large $p$ as
$p^{-\frac32+ m}$ and  $p^{-\frac32-m}$. Curiously, the case
$\alpha=\frac14$ is more problematic and has to be treated separately.}
%spanned by certain vectors that behave for large $p$ as
%$p^{-\frac32+ m}$ and  $p^{-\frac32-m}$.

 Finally, we define
self-adjoint
realizations of Bessel operators by selecting appropriate domains
larger than the minimal but smaller than the maximal domain.

The above construction of self-adjoint Bessel operators is
mathematically correct, but it sounds rather artificial. In the
remaining part of Sect. \ref{sec3} we describe ``physicists'
style''  construction, directly inspired by \cite{Glazek} and by the
literature on QFT in general. This construction involves the four
steps indicated in \eqref{subu}. Let us describe them more precisely.

The    counterterms will be constructed out of (unbounded and non-closable) operators $K_\mathrm{D}$ and
  $K_\mathrm{N}$:
  \begin{align}\label{popi}
    K_\mathrm{D}\psi(p)&:=p\int_0^\infty q\psi(q)\d q,\\
    K_\mathrm{N}\psi(p)&:=\int_0^\infty \psi(q)\d q,
  \end{align}
  % Let $\theta$ denote the Heaviside function.

Suppose that
$f_\Lambda$ and $g_\Lambda$ are solutions of  the differential equations
\begin{align}\label{diff1}
  \Lambda\frac{\mathrm{d}}{\mathrm{d}\Lambda} f_\Lambda 
  &=\Big(f_\Lambda+\frac14+\alpha\Big)^2-\alpha,\\\label{diff2}
    \Lambda\frac{\mathrm{d}}{\mathrm{d}\Lambda} g_\Lambda 
  &=-\Big(g_\Lambda{  +}\frac14{  +}\alpha\Big)^2 + \alpha. 
\end{align}
  {   Let  $p$ denote the momentum operator.}
Consider the operators
           \begin{align}
&  \theta(\Lambda-p) \Big( \Tilde{  L}_{\alpha,\mathrm{D}}+\frac{f_{\Lambda}}{\Lambda}
K_\mathrm{D}\Big)\theta(\Lambda-p)
  ,\label{mma1da}\\
 &  \theta(\Lambda-p)\Big( \Tilde{  L}_{\alpha,\mathrm{N}}
+\Lambda g_{\Lambda}
K_\mathrm{N}\Big)\theta(\Lambda-p).
\label{mma2da}
  \end{align}
Note that \eqref{mma1da} and \eqref{mma2da} are well-defined bounded self-adjoint
operators.
%On the other hand, the expressions in the brackets are ill
%defined. However, they are formally homogeneous of degree $2$.
{   We prove that for $\alpha<1$
%the operators \eqref{mma1da} and \eqref{mma2da}
they converge then in the strong
resolvent sense to self-adjoint realizations of the Bessel operator in
the momentum representation. This can be viewed as the main result
of our paper.}

According to ``physicists' style construction''
self-adjoint realizations of Bessel operators are
parametrized by real solutions to 
 the equations  \eqref{diff1} and
 \eqref{diff2}. Not surprisingly, these equations
 look very similar to  various equations for
running coupling constants used in QFT.
Note also
that  the right hand
sides of   \eqref{diff1} and
 \eqref{diff2} depend
analytically on $\alpha$ also at the ``phase transition''
$\alpha=0$. This phase transition is visible when we consider
real  solutions to these equations.

What is the take-home message of our paper? We do not insist that
Schr\"odinger operators should be always studied in the momentum
representation. However, we think that it is useful to know how the
world looks ``on 
the momentum side of the Fourier transform''. And it does
look 
quite differently from the position representation.

To our knowledge, mathematicians rarely   use the
momentum representation to study Schr\"odinger operators.
In particular, we have never seen the formulas 
\eqref{momen1a} and \eqref{momen2a} in mathematics papers. In physics
papers, the 
momentum representation seems to be more common. For instance, the formula
\eqref{momen1a} can be found in \cite{Glazek} (in the 3-dimensional
context).

The original purpose of our paper was to present the ideas of 
\cite{Glazek} in clear and rigorous terms. We think that these ideas 
are important and instructive. One can try to formulate them as
follows.
In renormalizable Quantum Field Theories
most terms in the Lagrangian have the same homogeneity wrt the
scaling.  E.g in  the Standard Model only the Higgs mass term has a
different scaling dimension. Bessel operators are also almost
homogeneous, with scaling invariance broken only by the boundary
condition at $0$. This homogeneity is related to the fact that the naive
formulations of both QFT and Bessel operators
in the momentum representation are highly singular.
Computations in  QFT usually involve
the four steps described in   \eqref{subu}:
 imposing a momentum cutoff $\Lambda$,
 adding counterterms,   solving the
differential equation for the coupling constant and taking the
limit $\Lambda\to\infty$.
Our paper   shows that the same steps have to
be followed in a much simpler context of Bessel operators. These steps
are not  restricted to the  difficult and complicated formalism
of QFT. They are  typical of the momentum approach.

{   Equations \eqref{diff1} and \eqref{diff2} have a
  one-parameter family of solutions for any
  $\alpha\in\mathbb{R}$. However, Bessel operators have a one-parameter
  family of self-adjoint realizations only for $\alpha<1$. In the
  language of QFT,
   ``nontrivial renormalized theories'' exist only for $\alpha<1$. For
  $\alpha\geq1$ the "flows of operators'' \eqref{mma1da} and
  \eqref{mma2da}  do not  have limits in the sense of
  the Hilbert space $L^2(\mathbb{R}_+)$, apart from  
   the unique self-adjoint realizalion of
  $L_\alpha$. Hence for $\alpha\geq1$, as Wilson suggested,
  ``one can never
  remove the cutoff completely in a nontrivial theory''.

  In particular, this is the case for the borderline value $\alpha=1$.
  The Bessel operator $L_1$ is the radial part of the Laplacian in 4
  dimensions, see \eqref{alpha}. 
 We believe that it is not a coincidence that  our spacetime  also has 4 dimensions.
}

%Note in parentheses that  Bessel operators are in some sense
%``too good'' to illustrate all ideas of the  Wilsonian
%renormalization. In fact, one can completely remove the cutoff and
%obtain a  ``renormalized theory'', something that may  be impossible
%in QFT, as Wilson suggested.  This ``renormalized theory'' in the case
%of Bessel operators is one of its self-adjoint  realizations.

\section{Position approach}\label{sec1}
%\init

\subsection{Extensions of Hermitian operators}
Let us recall basic concepts of the  theory of self-adjoint extensions
of Hermitian operators (often  called  symmetric operators). A comprehensive treatment of this topic can be found in \cite{ReedSimon}.

Let $A$ and $B$ be operators with domains $\mathcal{D}(A)$ and
$\mathcal{D}(B)$ respectively. We say that $A$ is {\em contained in} $B$ if $\mathcal{D}(A) \subseteq \mathcal{D}(B)$ and $A = B$ on $\mathcal{D}(A)$. We then write
 $A\subseteq B$.

An operator $A$ is {\em Hermitian} if for all $x,y \in \mathcal{D}(A)$
\begin{equation}
  (Ax|y) = (x|Ay),
\end{equation}
where $(\cdot| \cdot)$ is the scalar product.
An operator $A$ is self-adjoint if $A^*=A$, where the star denotes the Hermitian adjoint.

\subsection{Self-adjoint realizations of $L_\alpha$}
{ Recall that operators given by \eqref{eq-L-m-} with specified domains are called realizations of $L_\alpha$.} Following  \cite{Derezinski2} (see Sec. 4 and
App. A therein for details), we first introduce the {\em maximal}
and {\em minimal realizations} of $L_\alpha$. The maximal realization, denoted
 $L^{\mathrm{max}}_\alpha$, has the domain
\begin{equation}
    \mathcal{D}(L^{\mathrm{max}}_\alpha) = \big{\{} f \in L^2 (\mathbb{R}_+) \: | \: L_\alpha f \in L^2 (\mathbb{R}_+) \big{\}}.
\end{equation}
and the
minimal realization of $L_\alpha$, denoted $L_\alpha^{\min}$, is the closure of the restriction of $L_\alpha$ to $C^\infty_\mathrm{c}(\mathbb{R}_+)$. Obviously,  $L^{\mathrm{min}}_\alpha \subset L^{\mathrm{max}}_\alpha$. Moreover, one can show that
\begin{equation}
    \big{(}L^{\mathrm{min}}_\alpha\big{)}^* = L^{\mathrm{max}}_\alpha.
\end{equation}
Hence, $L^{\mathrm{min}}_\alpha$ is Hermitian.

%{ Recall that} functions in
% $C^\infty_\mathrm{c}(\mathbb{R}_+)$ are supported away from $0$.

One can show that for $\alpha \geq 1$
      $L_\alpha^{\max}=L_\alpha^{\min}$ is self-adjoint{  \cite{Derezinski2}}. In what follows it will be denoted $H_m$, where  $m:=\sqrt{\alpha}$ is the positive square root of $\alpha$.

     For  $\alpha < 1$
the domain of
$L_\alpha^{\min}$ is strictly smaller than that of $L_\alpha^{\max}$
and neither operator is self-adjoint. Therefore, both operators are of little use in physical applications: e.g. eigenvalues of  $L_\alpha^{\max}$ cover almost the whole complex plane and  corresponding eigenvectors are not mutually orthogonal.

Self-adjoint realisations of $L_\alpha$  are self-adjoint operators $H$
such that
\begin{equation}
 L^{\mathrm{min}}_\alpha \subset  H=H^*
  \subset L^{\mathrm{max}}_\alpha.
\end{equation}
They are constructed as follows. Firstly, one finds solutions of $L_\alpha \psi = 0$. These are of the form
\begin{equation}
    \psi(x)=\left\{
                \begin{array}{ll}
                  a\cdot x^{1/2+m} + b\cdot x^{1/2 - m}, & m \neq 0;\\
                  a\cdot \sqrt{x} + b\cdot \sqrt{x} \ln x, & m = 0.
                \end{array}
              \right.
    \label{eq-Hpsi=0}
\end{equation}
Here, $m:=\sqrt{\alpha}$. Note that
$m \in \mathbb{R}$ or $m\in \mathrm{i}\mathbb{R}$ is defined up to a sign.

For $\alpha\geq1$ only $ x^{1/2+m}$ with $m$ positive is square
integrable near zero. This is the key element of the proof of the
essential self-adjointness of  $H_m$ on $C^\infty_\mathrm{c}(\mathbb{R}_+)$.

For $\alpha<1$ all the solutions (\ref{eq-Hpsi=0}) are
square integrable {   on $]0,\varepsilon[$ for any
  $\varepsilon>0$.  Let $\kappa,\nu$ be complex numbers. Let $\xi\in
C_\mathrm{c}^\infty[0,\infty[$ and $\xi=1$ near $0$.
The following spaces lie between $\cD(L_\alpha^{\min})$ and 
$\cD(L_\alpha^{\max})$ and do not depend on the choice of $\xi$:
\begin{equation}
\begin{aligned}
&\mathcal{D}(H_{m,\kappa}) = \{ f \in
\mathcal{D}(L^{\mathrm{max}}_\alpha ) \quad | \quad \text{for some} \:  c \in \mathbb{C} \quad f(x) - c\xi(x)(\kappa x^{1/2 -m} + x^{1/2 + m}) \in \mathcal{D}(L^{\mathrm{min}}_\alpha )  \},\\
&\mathcal{D}(H_{m,\infty}) = \{ f \in
\mathcal{D}(L^{\mathrm{max}}_\alpha ) \quad | \quad \text{for some} \:  c \in \mathbb{C} \quad f(x) - c\xi(x) x^{1/2 -m} \in \mathcal{D}(L^{\mathrm{min}}_\alpha )  \},\\
&\mathcal{D}(H_{0}^\nu) = \{ f \in \mathcal{D}(L^{\mathrm{max}}_{0})
\quad | \quad \text{for some} \:  c \in \mathbb{C} \quad f(x) - c\xi(x)(x^{1/2}\ln x + \nu x^{1/2}) \in \mathcal{D}(L^{\mathrm{min}}_{0})  \}, \\
&\mathcal{D}(H_{0}^\infty) = \{ f \in
\mathcal{D}(L^{\mathrm{max}}_{0}) \quad | \quad \text{for some} \: c \in \mathbb{C} \quad f(x) - c\xi(x) x^{1/2} \in \mathcal{D}(L^{\mathrm{min}}_{0})  \}.
\end{aligned}
\label{eq-domains}
\end{equation}}
They define realizations of $L_\alpha$ denoted $H_{m,\kappa}$ and $H_0^\nu$.
The above definition implies immediately that
$H_{m,\kappa} = H_{-m, \kappa^{-1}}$.
Moreover, Hermitian adjoints of these operators are
$(H_{m,\kappa})^* = H_{\bar{m},\bar{\kappa}}$ and $(H_0^\nu)^* = H^{\bar{\nu}}_0$ (with the convention $\bar{\infty} = \infty$). \\[0.2cm]
To prove it, let us consider functions  $f\in\mathcal{D}(H_{m,\kappa})$, $g\in\mathcal{D}(H_{\bar{m},\bar{\kappa}})$. One has
$$
(H_{m,\kappa}f| g) - (f|H_{\bar{m},\bar{\kappa}}g) = \lim_{x\rightarrow 0}\Big{(}\bar{f}(x) \partial_x g(x) - g(x) \partial_x \bar{f}(x)\Big{)}.
$$
By analysis of behaviour of $f$ and $g$ near $x=0$ we conclude that the Wronskian at $0$ vanishes. Calculations for $H^{\nu}_0$ are analogous.\\

\noindent Thus,    self-adjoint extensions of $L^{\mathrm{min}}_{m^2}$ with $m^2<1$ fall into the following 3 categories:
\begin{itemize}
    \item $H_{m,\kappa}$, with $m\in ]0,1[$ and $\kappa \in \mathbb{R}\cup\{\infty\}$,
    \item $H_{m,\kappa}$, with  $m \in \mathrm{i}\mathbb{R}_+$ and $|\kappa|=1$,
    \item $H_0^\nu$ with  $\nu \in \mathbb{R}\cup\{\infty\}$.
\end{itemize}
{  In the second line, $|\kappa|=1$ is a consequence of
  $H_{m,\kappa} = H_{-m, \kappa^{-1}}$ and  $(H_{m,\kappa})^* =
  H_{\bar{m},\bar{\kappa}}$.}

Note that $H_{\frac12,0}=H_{-\frac12,\infty}$ and
  $H_{-\frac12,0}=H_{\frac12,\infty}$
  are  the Laplacians with the {\em Dirichlet}, resp. {\em Neumann  boundary
  condition at} $0$. They will be often denoted $H_\mathrm{D}$,
  resp. $H_\mathrm{N}$.

\subsection{Point spectra}

\noindent After this discussion, we are able to identify the point
spectra of the above Hamiltonians, following e.g. \cite{Derezinski1}. \\[0.2cm]
$\bullet \quad$ For $m\in]0,1[
    $ and $\kappa \in \mathbb{R}\cup\{\infty\}$:\\
\begin{equation}
\begin{aligned}
  &\sigma_\p (H_{m,\kappa}) = \bigg{\{} -4\Big{(}\frac{\kappa\Gamma(-m)}{\Gamma(m)}\Big{)}^{-1/m} \bigg{\}} \quad \mathrm{for} \quad \kappa \in]-\infty,0[
  ,\\
&\sigma_\p (H_{m,\kappa}) = \emptyset \quad \mathrm{for} \quad \kappa \in [0,\infty].
\label{eq-E-weak-alpha}
\end{aligned}
\end{equation}
$\bullet \quad$ For $m=\mathrm{i}m_I\in \mathrm{i}\mathbb{R}_+$ and $|\kappa| = 1$:
\begin{equation}
\sigma_\p (H_{\mathrm{i}m_I,\kappa}) = \bigg{\{} -4\exp \bigg{(}\frac{\arg\Big{(} \frac{\kappa\Gamma(-\mathrm{i}m_I)}{\Gamma(\mathrm{i}m_I)}\Big{)}+2\pi n}{m_I} \bigg{)} \quad | \quad n\in\mathbb{Z} \bigg{\}}.
\label{eq-E-strong-alpha}
\end{equation}
It implies that for $\alpha < 0$ the point spectrum of self-adjoint extensions of $L^{\mathrm{min}}_\alpha$ has infinitely many elements with accumulation points at $0$ and $-\infty$.\\[0.2cm]
$\bullet \quad$ For $\nu \in \mathbb{R} \cup \{ \infty \}$
\begin{equation}
\begin{aligned}
&\sigma_\p (H_0^\nu) = \{ -4\e^{2(\nu-\gamma)} \} \quad \nu \in \mathbb{R},\\
&\sigma_\p (H_0^\infty) = \emptyset.
\end{aligned}
\label{eq-E-alpha-0}
\end{equation}
$\gamma$ denotes the Euler constant. In all cases,  bound-state solutions to the eigenvalue problem $H_{m,\kappa} \psi = -k^2 \psi$ or $H_0^\nu \psi = -k^2 \psi$ are of the form
\begin{equation}
    \psi(x) = \sqrt{kx} K_m (kx) \quad \mathrm{or} \quad \sqrt{kx} K_0 (kx),
    \label{eq-solutions-x}
\end{equation}
where $k>0$ and $K_m(z)$ is the MacDonald function  of order $m$ \cite{Watson}. The proof can be found e.g. in \cite{Derezinski1}, Sec. 5.2.

\subsection{Renormalization group}

Let us introduce the
{\em dilation (scaling) operator} $U_\tau$. It is a unitary transformation on $L^2(\mathbb{R}_+)$ acting on functions in the following way:
\begin{equation}\label{scaling}
    U_\tau f(x) = \e^{\tau/2} f(\e^\tau x).
\end{equation}
We say that an operator $B$ is {\em homogeneous of degree} $n$ if
$$
U_\tau B U_{-\tau} = \e^{n\tau} B.
$$
$L^{\mathrm{max}}_m$ and $L^{\mathrm{min}}_m$ are homogeneous of
degree $-2$. However, the realizations $H_{m,\kappa}$ are homogeneous
only for $\kappa = 0$ or $\kappa = \infty$.  {   Realizations
 $H_0^\nu$ are homogeneous only for $\nu = 
\infty$.} Moreover \cite{Derezinski1},
\begin{equation}
    \begin{aligned}
        & U_\tau H_{m,\kappa} U_{-\tau} = \e^{-2\tau} H_{m,\e^{-2m\tau}\kappa}\\
        & U_\tau H_0^\nu U_{-\tau} = \e^{-2\tau} H_0^{\nu+\tau}.
    \end{aligned}
\end{equation}
Indeed, if $f\in \mathcal{D}(H_{m,\kappa})$, then $U_\tau f\in \mathcal{D}(H_{m,\e^{-2m\tau}\kappa})$, and if $f\in \mathcal{D}(H_0^\nu)$, then $U_\tau f\in \mathcal{D}(H_0^{\nu+\tau})$.

For the purpose of this article,  the action of
$\mathbb{R}\ni\tau\mapsto R_\tau$ defined in (\ref{rg}) can be called
the {\em renormalization group}.
In the set of self-adjoint realizations of  $L_\alpha$
we have the following
behaviors of the renormalization group flow:
\begin{itemize}
\item For $\alpha\geq1$ the set is one point.
\item  for $0<\alpha<1$ there are two fixed points:
{    attractive $H_{m,0}$ and
repulsive  $H_{m,\infty}$. The former  is  the
  Friedrichs extension of $L_{m^2}^{\min}$ and
  the latter is  its Krein extension;  see \cite{Derezinski1,GraOri}.}
  (We choose $m$ as  the positive square root of $\alpha$). 
\item  for $\alpha = 0$ there is only one fixed point:
  $H^{\infty}_0$.
\item  for $\alpha < 0$ there are no fixed points and the renormalization group  generates a cyclic flow.
\end{itemize}
%Here is an illustration of these flows borrowed from \cite{Derezinski3},
%hopefully self-explanatory.
%The names of the various phases can be treated as jokes. 

\medskip

\begin{figure}[H]
\centering
\includegraphics[width = 0.7 \textwidth]{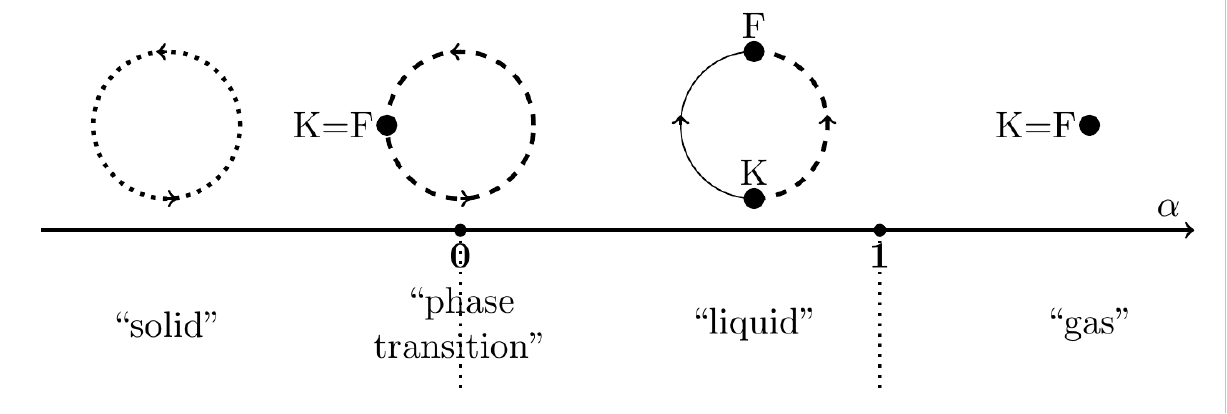}
\caption{{  
Schematic illustration of the renormalization group flow on the spaces
of self-adjoint realizations of Bessel operators. It is borrowed from \cite{Derezinski3}. Heavy dots represent
fixed points. The letters $K$ and $F$ correspond to the Krein and Friedrichs extensions 
of $L_\alpha^{\min}$. Dashed lines represent realizations with a single
bound state. Dotted line represents an infinite number of bound
states. The arrow indicates the direction of the flow.}}\label{figure}\end{figure}

{   The names of the various phases can be  treated as
jokes. One can also try to justify them as follows. For $\alpha>1$ 
the space of realizations is trivial. This  is  the simplest
phase---hence the name ''gas''.
For $\alpha<0$ the continuous scaling symmetry is broken to a discrete
subgroup, as in crystals---this justifies the name ``solid''.
The intermediate situation, $0<\alpha<1$ is consequently called the ``liquid phase''.
}

\bigskip

\section{``Wilsonian  approach'' to inverse square potential}
\label{sec2}
% \init

  In this section we give a partly heuristic
 theory of  Bessel operators in the spirit of the Wilsonian
renormalization. We will mostly follow the 
presentation of \cite{Glazek}, preserving to a large extent its style
and language. In the next section we give a rigorous description of
all steps of this section.
%In particular, we use the 3-dimensional formulation---as described in Subsect.
%\ref{Generalisation to the 3-dimensional problem}, it is equivalent to the 1-dimensional case. Analysis of this problem in the momentum representation can be found also in \cite{vanKolck, Long, Hammer, Braaten}.

For a  description of the Wilsonian renormalization procedure applied
to quantum field theory the reader can consult \cite{Wilson1, Wilson2, Wilson3}.

The main object of the analysis of
\cite{Glazek} is the formal expression
\begin{equation}
    \Tilde{L}_{\alpha,3} \psi(\vec{p}) = |\vec{p}|^2 \psi(\vec{p}) +\frac{1}{4\pi}\Big(\alpha-\frac{1}{4}\Big)\int d^3 \vec{q} \frac{1}{|\vec{p} - \vec{q}|}\psi(\vec{q}).
    \label{eq-momenta-1}
\end{equation}
%Formally, it is $L_{\alpha,3}$ of Eq. \eqref{eq-L-m3-laplacian}
%in the momentum representation. 
 Formally, it is the momentum representation of  
\begin{equation}
L_{\alpha,3}:=    -\Delta_3 + \Big( \alpha - \frac14 \Big) \frac{1}{r^2}, \label{eq-L-m3-laplacian}
\end{equation}
where $\Delta_3$ is the $3$-dimensional Laplacian and $r$ is the radial coordinate.

More precisely, if $\mathcal{F}f(\vec{p})$ is the Fourier transform of $f(\vec{x})$, then
\begin{equation}
  \Tilde{L}_{\alpha,3}:=    \mathcal{F} L_{\alpha,3} \mathcal{F}^{-1}.
\end{equation}
Let us note that the 3-dimensional formulation is equivalent to the
1-dimensional one.  Indeed, every $\psi\in L^2(\mathbb{R}^3)$ can be
written {   in spherical coordinates $r,\vartheta,\phi$} as
\[\psi=\sum_{l=0}^\infty\sum_{m=-l}^l Y_{l,m}(\vartheta,\phi)\frac1rf_{l,m}(r),\] where $Y_{l,m}$ are   spherical harmonics.  It follows that on  $l$ degree spherical harmonics 
$L_{\alpha,3}$ is equivalent to 
\[-\partial_r^2+\Big((l+1/2)^2+\alpha-\frac12\Big)\frac1{r^2}.\]

For further convenience let us denote $p := |\vec{p}|$. We will
restrict our attention {    to the  s-wave sector, that is, to
  spherically symmetric functions.
In the s-wave sector  we can write $\psi(\vec{p})
  = \psi(p)$ with $p:=|\vec p|$ and } the operator Eq. \eqref{eq-momenta-1} 
can be written in the following way (see Eqs. (3) and (4) in \cite{Glazek}):
\begin{equation}
    \Tilde{L}_\alpha \psi(p) = p^2 \psi(p) + \Big(\alpha-\frac{1}{4}\Big) \int_0^\infty \d q \: q^2 \Big{(} \theta(p-q) p^{-1} + \theta(q-p) q^{-1} \Big{)} \psi(q),
    \label{eq-L-tilde-s}
\end{equation}
where $\theta(x)$ denotes the Heaviside step function and we dropped $3$ from the subscript.

Unfortunately, \eqref{eq-L-tilde-s} does not define a Hermitian operator. To find a self-adjoint realisation, we introduce an ultraviolet cutoff $\Lambda$ and a counter-term $\frac{f_\Lambda}{\Lambda}$. Thus we consider a family of cutoff Hamiltonians
\begin{equation}
    \Tilde{H}_\alpha(\Lambda, f_\Lambda) \psi(p) \coloneqq \begin{cases} p^2 \psi(p) + \int_0^\Lambda \d q \: q^2 \Big{(} V(p,q) + \frac{f_\Lambda}{\Lambda} \Big{)} \psi(q), & \mathrm{for }\: p \leq \Lambda \\
    0, & \mathrm{for } \: p > \Lambda.
    \end{cases}
    \label{eq-H-Lambda}
\end{equation}
where 
\begin{equation}
    V(p,q) = \Big(\alpha-\frac{1}{4}\Big)\Big{(} \theta(p-q) p^{-1} + \theta(q-p) q^{-1} \Big{)}.
\end{equation}

{  
Let $\Tilde{U}$ denote the dilation in the momentum representation, that is
\begin{equation}
  \Tilde{U}(\tau)\mathcal{F}=\mathcal{F}U(\tau).
  \end{equation}
Note that in the momentum representation the dilation acts in the opposite way than in the position representation:
\begin{equation}
\Tilde{U}(\tau)\psi(p)=\e^{-\frac{3}{2}\tau}\psi(\e^{-\tau}p).
  \end{equation}
It is easy to see that
\begin{equation}
  \Tilde{U}(\tau)\Tilde{H}_\alpha(\Lambda,f_\Lambda)\Tilde{U}(-\tau)
  =\e^{-2\tau}\Tilde{H}_\alpha(\e^\tau\Lambda,f_\Lambda).
\end{equation}
Thus changing $\Lambda$ can be interpreted as the change of a scale.}

We would like to find out for what kind of dependence of $f_\Lambda$
on $\Lambda$ {   we can expect} the existence of a limit $H_\alpha(\Lambda,f_\Lambda)$
for $\Lambda\to\infty$ as a self-adjoint operator. To this end we assume that we have a fixed eigenfunction $\psi$ satisfying
\begin{equation}
  H_\alpha(\Lambda,f_\Lambda)\psi(p)=E\psi(p),\quad p<\Lambda.
  \end{equation}
Following the terminology of \cite{Glazek},
we will say that two Hamiltonians $\Tilde{H}_m(\Lambda, f_\Lambda)$ and $\Tilde{H}_\alpha(\Lambda', f_{\Lambda'})$ are equivalent if
$$
    \Tilde{H}_\alpha(\Lambda, f_\Lambda) \psi(p) =
    \Tilde{H}_\alpha(\Lambda', f_{\Lambda'}) \psi(p) \quad \mathrm{for} \: p<\mathrm{min}(\Lambda, \Lambda').
$$
Two Hamiltonians are equivalent if the function $\gamma(\Lambda) \equiv f_\Lambda / \Lambda$ satisfies the equation
\begin{equation}
    \frac{\mathrm{d}\gamma(\Lambda)}{\mathrm{d}\Lambda} = \Big{[} \gamma(\Lambda) - \frac{1/4 - \alpha}{\Lambda} \Big{]}^2.
    \label{eq-dgamma-dLambda}
\end{equation}
To prove it, we consider an infinitesimal reduction of the cut-off $\Lambda \rightarrow \Lambda - \d\Lambda$. Let $\psi$ be a solution of
\begin{equation}
\Tilde{H}_\alpha(\Lambda, f_\Lambda) \psi(p) = -k^2 \psi(p).
\label{eq-H-Lambda-psi-k2}
\end{equation}
By splitting the integral
$$
\int_0^\Lambda \d q \: q^2 \Big{(} V(p,q) + \gamma(\Lambda) \Big{)} \psi(q) \approx \int_0^{\Lambda-d\Lambda} \d q \: q^2 \Big{(} V(p,q) + \gamma(\Lambda) \Big{)} \psi(q) + \d\Lambda \Big{(} V(p,\Lambda) + \gamma(\Lambda) \Big{)}\psi({\Lambda}),
$$
and using the relation resulting from Eqs. \eqref{eq-H-Lambda} and \eqref{eq-H-Lambda-psi-k2}:
$$
    \psi(\Lambda) = -\frac{1}{k^2 + \Lambda^2} \int_0^\Lambda \d q\: q^2 \Big{(} (\alpha-\frac{1}{4}) \Lambda^{-1} + \gamma(\Lambda) \Big{)} \psi(q),
$$
assuming the large cut-off $k^2 \ll \Lambda^2$ we conclude that the Hamiltonian $\Tilde{H}_\alpha(\Lambda, f_\Lambda)$ is equivalent to $\Tilde{H}_\alpha(\Lambda-\d\Lambda, f_{\Lambda-\d\Lambda})$ such that
$$
\Tilde{H}_\alpha(\Lambda-\d\Lambda, f_{\Lambda-\d\Lambda}) \psi(p) \coloneqq p^2 \psi(p) + \int_0^{\Lambda-\d\Lambda} \d q \: q^2 \Big{(} V(p,q) + \gamma(\Lambda-\d\Lambda) \Big{)} \psi(q),
$$
where
\begin{equation}
    \gamma(\Lambda-\d\Lambda) = \gamma(\Lambda) - \d\Lambda \Big{[} \gamma(\Lambda) - \frac{1/4 - \alpha}{\Lambda} \Big{]}^2.
\end{equation}
Taking the limit $\d\Lambda \rightarrow 0$ we obtain the differential
equation \eqref{eq-dgamma-dLambda}. Its solutions are of the form
$\gamma(\Lambda) = \Lambda^{-1} f_\Lambda$. {   Let us analyze the flow
$\Lambda\to f_\Lambda$.
\begin{itemize}
    \item For  $\alpha<0$, we set $m_I:=\sqrt{-\alpha}$,
    \begin{equation}
        f_\Lambda = m_I \tan \bigg{(} \arctan \Big{(} \frac{f_{\Lambda_0} -m_I^2 + 1/4}{m_I} \Big{)} + m_I \log \frac{\Lambda}{\Lambda_0} \bigg{)} + m_I^2 - \frac{1}{4}.
        \label{eq-gamma-m-iR}
    \end{equation}
 There are no fixed points. Moreover, there is a discrete scaling symmetry: for any $n\in \mathbb{Z}$, Hamiltonians $\Tilde{H}(\Lambda, f_\Lambda)$ and $\Tilde{H}(\Lambda \cdot \exp(\pi n/m_I), f_\Lambda)$ are equivalent.
    \item For $\alpha = 0$,
    \begin{equation}
        f_\Lambda = \frac{f_{\Lambda_0} + 1/4}{1 - (f_{\Lambda_0}+1/4) \log \frac{\Lambda}{\Lambda_0}} - \frac{1}{4}.
        \label{eq-gamma-m-0}
      \end{equation}
      There is one fixed point corresponding to $f_{\Lambda_0} =
      -1/4$,
       for which Hamiltonians $\Tilde{H}_m(\Lambda, f_{\Lambda_0})$ are equivalent for all values of $\Lambda$.
            \item for $\alpha>0$, we set $m:=\sqrt{\alpha}>0$
    \begin{equation}
        f_\Lambda = -m \tanh \bigg{(}- \mathrm{atanh} \Big{(} \frac{f_{\Lambda_0} +m^2 + 1/4}{m} \Big{)} + m \log \frac{\Lambda}{\Lambda_0} \bigg{)} - m^2 - \frac{1}{4}.
        \label{eq-gamma-m-in-R}
      \end{equation}
 There are two fixed point: an attractive one     $f_{\Lambda_0}^+ =
 -\big(m+\frac12\big)^2$ and a repulsive one  $f_{\Lambda_0}^- =
 \big(m-\frac12\big)^2$,  for which Hamiltonians $\Tilde{H}_m(\Lambda, f^\pm_{\Lambda_0})$ are equivalent for all values of $\Lambda$.
\end{itemize}
}

%Let us now choose a fixed cut-off scale $\Lambda_0$ and analyze how these operators depend on the parameter $f_{\Lambda_0}$:
%\begin{itemize}
%\item For $0<\alpha<1$, $m=\sqrt{\alpha}$, we have $-|m|-m^2-1/4 \leq f_{\Lambda_0} \leq |m|-m^2-1/4$. For these extreme values, the atanh function becomes $\pm \infty$. For such an initial condition, the function $f_\Lambda$ is independent on $\Lambda$. It follows that in this case there are two fixed points: $f_{\Lambda_0}^{\pm} = \pm |m|-m^2-1/4$, for which Hamiltonians $\Tilde{H}_m(\Lambda, f^\pm_{\Lambda_0})$ are equivalent for all values of $\Lambda$.

%\item For $\alpha=0$ one has $f_{\Lambda_0} \in \mathbb{R} \cup \{ \infty \}$ and there is one fixed point corresponding to $f_{\Lambda_0} = -1/4$.

%\item For $\alpha<0$,  $m_I:=\sqrt{-\alpha}$ we have $f_{\Lambda_0} \in \mathbb{R} \cup \{ \infty \}$ and there are no fixed points. Moreover, there is a discrete scaling symmetry: for any $n\in \mathbb{Z}$, Hamiltonians $\Tilde{H}(\Lambda, f_\Lambda)$ and $\Tilde{H}(\Lambda \cdot \exp(\pi n/m_I), f_\Lambda)$ are equivalent.
%\end{itemize}

We thus reproduce the {   the three first pictures from
  Figure \ref{figure}.}
  %that we know from Sect. \ref{sec1}.

\section{Momentum approach}\label{sec3}
%\init
\subsection{Momentum approach to Schr\"odinger operators on $\mathbb{R}^d$}

Self-adjoint operators on $L^2(\mathbb{R}^d)$
of the form
\begin{equation}
  H=-\Delta+V(x)\end{equation}
are usually called {\em Schr\"odinger operators}.
They are typically studied by methods of the configuration space.
However, they can  also be approached from the momentum point of view.

{  
We will typically use $x\in\rr^d$ for the generic variable
in the position representation and $p\in\rr^d$ in the momentum representation.}
Let us recall the  two most common conventions for the Fourier
transformation:
\begin{align}\label{asin}
  \Tilde{f}(p)&:=\int \e^{-\mathrm{i} xp}f(x)\mathrm{d} x,\\
  \cF f(p)&:=(2\pi)^{-\frac{d}{2}}\int \e^{-\mathrm{i} xp}f(x)\mathrm{d} x=
  (2\pi)^{-\frac{d}{2}}\Tilde{f}(p).
\end{align}
Note that $\cF$ is unitary. Let $\Tilde{H}$ denote the operator $H$ in the momentum representation, that is
\begin{equation}\Tilde{H}:=\cF H\cF^{-1}.\end{equation}
If the potential $V$ is well-behaved, then $\Tilde{H}$ can be written as
\begin{equation}
  \Tilde{H}\psi(p)=p^2\psi(p)+(2\pi)^{-d}\int\Tilde{ V}(p-q)\psi(q)\mathrm{d} q.\label{momen}\end{equation}

\subsection{Momentum approach to Schr\"odinger operators on $\mathbb{R}_+$}

Often we consider Schr\"odinger operators on a subset of $\mathbb{R}^d$.
Then  the representation (\ref{momen}) becomes problematic, because
it needs to take into account the boundary conditions. Besides, it is not obvious what should replace the Fourier transformation.

A case of particular interest is $\mathbb{R}_+$. In this case we
have two distinguished realizations of the 1-dimensional Laplacian:
the {\em Dirichlet Laplacian} $\Delta_\mathrm{D}$ and the {\em Neumann Laplacian} $\Delta_\mathrm{N}$.
Instead of the Fourier transformation it is natural to use the {\em
  sine transformation} $\cF_{\mathrm{D}}$ or the {\em cosine transformation} $\cF_{\mathrm{N}}$:
\begin{align}\label{fd}
  (  \cF_\mathrm{D}f)(p)&:=\sqrt{\frac{2}{\pi}}\int_0^\infty\sin(px) f(x)\mathrm{d}x,\\\label{fn}
    (  \cF_\mathrm{N}f)(p)&:=\sqrt{\frac{2}{\pi}}\int_0^\infty\cos(px) f(x)\mathrm{d}x.
\end{align}
Both are involutions, that is,
\begin{equation}
  \cF_\mathrm{D}^2\psi=  \cF_\mathrm{N}^2\psi=\psi,
\end{equation}
they are  unitary and diagonalize the
Dirichlet/Neumann Laplacian:
\begin{align}
  \cF_\mathrm{D}(-\Delta_\mathrm{D})  \cF_\mathrm{D}\psi(p)&=p^2\psi(p), \label{Eq-why-D}\\
  \cF_\mathrm{N}(-\Delta_\mathrm{N})  \cF_\mathrm{N}\psi(p)&=p^2\psi(p). \label{Eq-why-N}
\end{align}

Consider now a Schr\"odinger operator on the half-line with a potential $V$.
Let us first assume that $V$ is sufficiently regular, say, $V\in L^1(\mathbb{R}_+)$. Then one can impose the Dirichlet or Neumann boundary conditions at $0$, so that one has two Schr\"odinger operators
\begin{align}
  H_\mathrm{D}&:=-\Delta_\mathrm{D}+V(x),\\
  H_\mathrm{N}&:=-\Delta_\mathrm{N}+V(x).
  \end{align}
 {  As suggested by Eqs. \eqref{Eq-why-D}-\eqref{Eq-why-N}}, to obtain their momentum versions it is natural to apply the sine transform to the first and the cosine transform to the second Hamiltonian:
\begin{align}
 \label{momi1} \Tilde{  H}_\mathrm{D}&:= \cF_\mathrm{D} H_\mathrm{D} \cF_\mathrm{D}
  ,\\\label{momi2}
    \Tilde{  H}_\mathrm{N}&:= \cF_\mathrm{N} H_\mathrm{N} \cF_\mathrm{N}.
  \end{align}

Let us extend the potential $V$ to an even function on the whole $\mathbb{R}$, so that $V(-x)=V(x)$. Let $\Tilde{V}(p)$ be the Fourier transform of $V$, as in (\ref{asin}).
Then \eqref{momi1} and \eqref{momi2} are given by the following
expressions, which easily follow from
(\ref{momen}) and \eqref{restri}:
\begin{align}
  \Tilde{H}_\mathrm{D}\psi(p)=p^2\psi(p)+\frac{1}{2\pi}\int_0^\infty(\Tilde{ V}(p-q)-\Tilde{ V}(p+q)\big)\psi(q)\mathrm{d} q,\label{momen1}\\
    \Tilde{H}_\mathrm{N}\psi(p)=p^2\psi(p)+\frac{1}{2\pi}\int_0^\infty(\Tilde{ V}(p-q)+\Tilde{ V}(p+q)\big)\psi(q)\mathrm{d} q.\label{momen2}
\end{align}

\subsection{Minimal realization in the momentum approach}

  A priori it is not obvious which transformation is more appropriate for the
  Bessel operator: the sine or the cosine transformation. Let us consider both, setting
\begin{align}
  \Tilde{  L}_{\alpha,\mathrm{D}}&:= \cF_\mathrm{D} L_\alpha\cF_\mathrm{D}
  ,\\
    \Tilde{  L}_{\alpha,\mathrm{N}}&:= \cF_\mathrm{N} L_\alpha \cF_\mathrm{N},
\end{align}
(where, as before, $L_\alpha$ is treated as a formal expression).

Then, using $-\pi |p|$ as the Fourier transform of $\frac{1}{x^2}$, 
(\ref{momen1}) and (\ref{momen2}) yield
  \begin{align}
    \Tilde{  L}_{\alpha,\mathrm{D}}\psi(p)&
    =
    p^2\psi(p)+\Big(\alpha-\frac14\Big)\int_0^\infty\big(\theta(p-q)q
    +\theta(q-p)p\big)\psi(q) \mathrm{d}q
  ,\label{mm1}\\
    \Tilde{  L}_{\alpha,\mathrm{N}}\psi(p)&
    =
    p^2\psi(p)-\Big(\alpha-\frac14\Big)\int_0^\infty\big(\theta(p-q)p
    +\theta(q-p)q\big)\psi(q)\mathrm{d}q.\label{mm2}
  \end{align}

So far, we have treated 
$    \Tilde{  L}_{\alpha,\mathrm{D}}$ and $    \Tilde{  L}_{\alpha,\mathrm{N}}$
as formal expressions. We would like to make them well defined, first
in the ``minimal sense''.

 In the position representation, an easy way to make
    the expression $L_\alpha$ well defined was to restrict its domain to
    $C_\mathrm{c}^\infty(\rr_+)$, which after the closure led to the
    minimal realisation. Of course, it was not necessary to demand
    that the support is away from zero, however  functions
    in the domain should have vanished
    of an appropriate order at zero.    In the momentum representation
 it is useful to restrict the domain to rapidly decaying
 functions---in the position representation this guarantees the
 smoothness.
On the position side,  this also guarantees that even/odd derivatives
at zero
 vanish in the Dirichlet, resp. Neumann case, see \eqref{formu2}. This does not
 suffice---in addition, we need to impose conditions that on the position
 side yield vanishing first/zeroth derivative.
 Thus, we introduce the spaces  (see Appx \ref{appx}):
\begin{equation}\label{amm1}
 L_j^{2,\infty}(\rr_+)  :=\left\{\psi\in L^{2,\infty}(\rr_+)\ |\
\int_0^\infty|\psi(p)|^2p^n\d p<\infty,\quad n=0,1,\dots;\quad    \int_0^\infty\psi(p)p^j\mathrm{d} p =0 
\right\},\quad j=0,1.
\end{equation}

\begin{theorem}\label{th-L-min}
\begin{enumerate}\item
  The operator
  $    \Tilde{  L}_{\alpha,\mathrm{D}}$
given by the
formula (\ref{mm1}) restricted to $ L_1^{2,\infty}(\rr_+)$
is well defined and closable. Let us denote its
closure by   $    \Tilde{  L}_{\alpha,\mathrm{D}}^{\min}$. We then have
\begin{align}\label{mnn1}
    \Tilde{  L}_{\alpha,\mathrm{D}}^{\min}&:= \cF_\mathrm{D} L_\alpha^{\min} \cF_\mathrm{D}.
\end{align}
\item 
The operator
    $    \Tilde{  L}_{\alpha,\mathrm{N}}$
    given by the formula  (\ref{mm2}) restricted to $ L_0^{2,\infty}(\rr_+)$
is well defined and closable. Let us denote its
closure by
  $    \Tilde{  L}_{\alpha,\mathrm{N}}^{\min}$. Then we have
\begin{align}\label{mnn2}
    \Tilde{  L}_{\alpha,\mathrm{N}}^{\min}&:= \cF_\mathrm{N} L_\alpha^{\min} \cF_\mathrm{N}.
\end{align}
\end{enumerate}
\label{theorem-L-min}  \end{theorem}

\noindent{\it Proof.} Consider first the Dirichlet case.
 Let $\psi\in L_1^{2,\infty}(\rr_+)$ and $x>0$. Consider
\begin{align}
  \phi(p)&:=\int_0^\infty \big(\theta(p-q)q+\theta(q-p)p\big)\psi(q) \mathrm{d}q =\int_p^\infty(p-q)\psi(q)\mathrm{d}q,
           \end{align}
{ where in the last transition we used the property of Eq. \eqref{amm1}.}
           Clearly, $\phi(0)=0$ and $\lim\limits_{p\to\infty}\phi(p)=0$. Moreover,
           \begin{equation}
             \phi'(p)=\int_p^\infty \psi(q)\mathrm{d} q\end{equation}
           and $\lim\limits_{p\to\infty}\phi'(p)=0$. We can integrate twice
           obtaining
           \begin{align}
             \int_0^\infty\sin(px)\phi(p)\mathrm{d}
             p&=\frac1x\int_0^\infty\cos(px)\phi'(p)\mathrm{d}p\\
              &=\frac1{x^2}\int_0^\infty\sin(px)\psi(p)\mathrm{d}p. 
           \end{align}

By \eqref{formu3},
\begin{align}
             \int_0^\infty\sin(px)p^2\phi(p)\mathrm{d}
             p&=-\partial_x^2\int_0^\infty\sin(px)\phi(p)\mathrm{d}p.           \end{align}
           
           Therefore,
           \begin{equation}
             \cF_{\mathrm{D}}\tilde L_{\alpha,\mathrm{D}}\psi=
 L_{\alpha}                          \cF_{\mathrm{D}}\psi.\end{equation}
Thus $  \cF_{\mathrm{D}}\tilde L_{\alpha,\mathrm{D}}\cF_{\mathrm{D}}$
coincides with $L_\alpha$ on 
$\cF_{\mathrm{D}}L_1^{2,\infty}(\rr_+)$.

Suppose now that $\phi= \cF_{\mathrm{D}}\psi$ with $\psi\in
L_1^{2,\infty}(\rr_+)$. We have $\phi^{(n)}(0)=0$, $n=0,2,\dots$
(by \eqref{formu2}, because $\psi\in
L^{2,\infty}(\rr_+)$). Moreover, $\phi'(0)=0$ (by \eqref{formu4}, because
$\int\limits_0^\infty p\psi(p)\mathrm{d}p=0$).
Hence $\phi(x)=O(x^3)$.

Let $\chi,\xi\in C^\infty(\rr_+)$, $\chi=1$,
$\xi=0$ near $0$ and $\chi=0$, $\xi=1$ near $\infty$.
Set $\phi_\epsilon(x):=\chi(x/\epsilon)\xi(x\epsilon)\psi(x)$.
Then
we easily show that
$\|L_\alpha(\phi-\phi_\epsilon)\|^2+\|
\phi-\phi_\epsilon\|^2\to0$ {    as $\epsilon \rightarrow 0$.} Clearly, $\phi_\epsilon\in
C_\mathrm{c}^\infty(\rr_+)$. Therefore $
\cF_{\mathrm{D}}L_1^{2,\infty}(\rr_+)\subset
\cD(L_\alpha^{\min})$.

Clearly, $  C_{\mathrm{c}}^\infty ( \mathbb{R}_+ )\subset\cF_{\mathrm{D}}L_1^{2,\infty}(\rr_+)$.
Therefore, $\cF_{\mathrm{D}}L_1^{2,\infty}(\rr_+)$ is dense in $\cD(L_\alpha^{\min})$.

This proves that the closure of $  \cF_{\mathrm{D}}\tilde
L_{\alpha,\mathrm{D}}\cF_{\mathrm{D}}$
restricted to $\cF_{\mathrm{D}}L_1^{2,\infty}(\rr_+)$ 
coincides with $L_\alpha^{\min}$.

Consider next the Neumann case.
 Let $\psi\in L_0^{2,\infty}(\rr_+)$. Consider
\begin{align}
  \phi(p)&:=\int_0^\infty\mathrm{d}q\big(\theta(p-q)p+\theta(q-p)q\big)\psi(q)=\int_p^\infty(q-p)\psi(q)\mathrm{d}q.
           \end{align}
           Clearly, $\lim\limits_{p\to\infty}\phi(p)=0$. Moreover,
           \begin{equation}
             \phi'(p)=\int_0^p \psi(q)\mathrm{d} q\end{equation}
           and $\phi'(p)=0$, $\lim\limits_{p\to\infty}\phi'(p)=0$. We can integrate twice
           obtaining
           \begin{align}
             \int_0^\infty\cos(px)\phi(p)\mathrm{d}
             p&=-\frac1x\int_0^\infty\sin(px)\phi'(p)\mathrm{d}p\\
              &={ -}\frac1{x^2}\int_0^\infty\cos(px)\psi(p)\mathrm{d}p.
           \end{align}

Clearly, 
\begin{align}
             \int_0^\infty\cos(px)p^2\phi(p)\mathrm{d}
             p&=-\partial_x^2\int_0^\infty\cos(px)\phi(p)\mathrm{d}p.           \end{align}           

Therefore
           \begin{equation}
             \cF_{\mathrm{N}}\tilde L_{\alpha,\mathrm{N}}\psi=
 L_{\alpha}                          \cF_{\mathrm{N}}\psi.\end{equation}
Thus $  \cF_{\mathrm{N}}\tilde L_{\alpha,\mathrm{N}}\cF_{\mathrm{N}}$
coincides with $L_\alpha$ on 
$\cF_{\mathrm{N}}L_0^{2,\infty}(\rr_+)$.

Suppose now that $\phi= \cF_{\mathrm{N}}\psi$ with $\psi\in
L_0^{2,\infty}(\rr_+)$. We have $\phi^{(n)}(0)=0$, $n=1,3,\dots$
(because $\psi\in
L^{2,\infty}(\rr_+)$). Moreover, $\phi(0)=0$ (because
$\int\limits_0^\infty \psi(p)\mathrm{d}p=0$).
Hence $\phi(x)=O(x^2)$.

The remaining arguments are the same as in the Dirichlet case.
\qed
  
  \subsection{Maximal realization in the momentum approach}\label{subsec-4-4}

  {   In this subsection we present a construction of the
    maximal operator
    $L_\alpha^{\max}$ for  $\alpha<1$. The construction involves two
    steps:
    \begin{enumerate}
    \item
      the  choice of two vectors that together with 
      $\cD(    L_\alpha^{\min})$ span $\cD(    L_\alpha^{\max})$;
    \item the action of $  L_\alpha^{\max}$ on these two vectors.\end{enumerate}

We will first describe the construction for the generic case
$\alpha\neq\frac14$.
Let $\alpha<1$,
% $\alpha\neq\frac14$
and  
$\alpha=m^2$. As we remember, if $\alpha\neq0$, elements of $ \cD ( L^{\mathrm{max}}_{\alpha} )$
behave  like $c_+ x^{1/2+m} + c_- x^{1/2-m}$ as $x\rightarrow 0$.
By
\eqref{eq-sin-tr}  and \eqref{eq-cos-tr}, 
the sine and cosine transforms of $x^{\frac12\pm m}$ are proportional to 
$p^{-\frac32\mp m}$.
This suggests to enlarge the domain of 
the  minimal operators by functions that behave like 
$p^{-\frac32\mp m}$ as $p\rightarrow \infty$.
However, we cannot simply take $p^{-\frac32\mp m}$, because they are not
square integrable near zero. This complicates the whole story.
(For $\alpha=0$, additionally,  we need to modify all this using the logarithm.)}

More precisely, let us fix
$\psi_{\pm m,\mathrm{D}}$, $\psi_{\pm m,\mathrm{N}}$, $\psi_{\log,\mathrm{D}}$ and $\psi_{\log,\mathrm{N}}$
be functions in $L^2(\rr_+)$ satisfying the following condition:
 there exists $\Lambda_0>0$ such that
\begin{align}\label{pqw1}
  p>\Lambda_0\ \Rightarrow\quad
  \psi_{\pm m,\mathrm{D}}(p)=p^{-\frac32\mp m},&\qquad
  \int_0^{\Lambda_0} \psi_{\pm
  m,\mathrm{D}}(q)q\mathrm{d}q
                                             =\frac{\Lambda_0^{\frac12\mp
                                              m}}{\frac12\mp m}, \quad {  m\neq \pm\frac12},\\
\label{pqw1log}
  p>\Lambda_0\ \Rightarrow\quad
  \psi_{\log,\mathrm{D}}(p)=p^{-\frac32}\log(p),&\qquad
  \int_0^{\Lambda_0} \psi_{\log,\mathrm{D}}(q)q\mathrm{d}q
                                             =-4 \Lambda_0^{\frac12} + 2\Lambda_0^{\frac12}\log \Lambda_0 ,\\
  \label{pqw2}
  p>\Lambda_0\ \Rightarrow\quad
  \psi_{\pm m,\mathrm{N}}(p)=p^{-\frac32\mp m},&\qquad
                             \int_0^{\Lambda_0} \psi_{\pm
  m,\mathrm{N}}(q)\mathrm{d}q
=\frac{\Lambda_0^{-\frac12\mp m}}{-\frac12\mp m},  \quad {  m\neq \mp\frac12},\\
  \label{pqw2log}
  p>\Lambda_0\ \Rightarrow\quad
  \psi_{\log,\mathrm{N}}(p)=p^{-\frac32}\log(p),&\qquad
                             \int_0^{\Lambda_0} \psi_{\log,\mathrm{N}}(q)\mathrm{d}q
=-4 \Lambda_0^{-\frac12} -2\Lambda_0^{-\frac12}\log\Lambda_0.                                                      \end{align}
Note that \eqref{pqw1} and \eqref{pqw2} imply
\begin{align}\label{pqw1-}
                                                  \Lambda>\Lambda_0\ \Rightarrow&
  \int_0^\Lambda \psi_{\pm
  m,\mathrm{D}}(q)q\mathrm{d}q
                                             =\frac{\Lambda^{\frac12\mp
                                                                                  m}}{\frac12\mp m},
   \quad {  m\neq \pm\frac12},\\
\label{pqw1log-}
                                                  \Lambda>\Lambda_0\ \Rightarrow&
  \int_0^\Lambda \psi_{\log,\mathrm{D}}(q)q\mathrm{d}q
                                             =-4 \Lambda^{\frac12} + 2\Lambda^{\frac12}\log \Lambda,\\
  \label{pqw2-}
                                               \Lambda>\Lambda_0\ \Rightarrow&
 \int_0^\Lambda \psi_{\pm
  m,\mathrm{N}}(q)\mathrm{d}q
=\frac{\Lambda^{-\frac12\mp m}}{-\frac12\mp m},  \quad {  m\neq \mp\frac12},\\
  \label{pqw2log-}
                                               \Lambda>\Lambda_0\ \Rightarrow&
 \int_0^\Lambda \psi_{\log,\mathrm{N}}(q)\mathrm{d}q
=-4 \Lambda^{-\frac12} -2\Lambda^{-\frac12}\log\Lambda.
                                                 \end{align}

We set
\begin{align}
  \cD
  (  \Tilde{L}_{\alpha,\mathrm{D}}^{\max})&:=\cD(\Tilde{L}_{\alpha,\mathrm{D}}^{\min})+
                                            \mathbb{C}\psi_{m,\mathrm{D}}+
 \mathbb{C}\psi_{-m,\mathrm{D}}, \quad {  \alpha\neq \frac14}\label{wna1}\\
\cD
  (  \Tilde{L}_{0,\mathrm{D}}^{\max})&:=\cD(\Tilde{L}_{0,\mathrm{D}}^{\min})+
                                            \mathbb{C}\psi_{0,\mathrm{D}}+
 \mathbb{C}\psi_{\log,\mathrm{D}}\label{wna10}\\
   \cD
  (
  \Tilde{L}_{\alpha,\mathrm{N}}^{\max})&:=\cD(\Tilde{L}_{\alpha,\mathrm{N}}^{\min})+
                                                        \mathbb{C}\psi_{m,\mathrm{N}}+
 \mathbb{C}\psi_{-m,\mathrm{N}}, \quad {  \alpha\neq \frac14}
,\label{wna2}\\
\cD
  (
  \Tilde{L}_{0,\mathrm{N}}^{\max})&:=\cD(\Tilde{L}_{0,\mathrm{N}}^{\min})+
                                                        \mathbb{C}\psi_{0,\mathrm{N}}+
 \mathbb{C}\psi_{\log,\mathrm{N}}
.\label{wna20}
\end{align}
Clearly, the difference of two functions
satisfying \eqref{pqw1}, \eqref{pqw1log}, \eqref{pqw2} or \eqref{pqw2log}, belongs to
$L_1^{2,\infty}(\rr_+)$, resp to $L_0^{2,\infty}(\rr_+)$. Therefore,  \eqref{wna1}, \eqref{wna2}, \eqref{wna10} and \eqref{wna20}
do not depend on the choices of $\psi_{\pm m,\mathrm{D}}$ and $\psi_{\pm m,\mathrm{N}}$.

The formulas (\ref{mm1}) and (\ref{mm2}) are in general ill defined on
\eqref{wna1}, \eqref{wna2}, \eqref{wna10} and \eqref{wna20}, because the integrals can be divergent at $\infty$.
In order to define the maximal operators in the momentum representation, we note that every $\psi$ in $ \cD  (  \Tilde{L}_{\alpha,\mathrm{D}}^{\max})$
or in $ \cD  (  \Tilde{L}_{\alpha,\mathrm{N}}^{\max})$ behaves like
\begin{equation}\psi(p) \sim \Tilde{c}_+ p^{-\frac32 - m} +
  \Tilde{c}_- p^{-\frac32 + m},\quad p\rightarrow
  \infty,\label{eq-asymptotic-p}\end{equation} 
\begin{equation}\text {   resp.}\qquad\psi(p) \sim \Tilde{c}_+ p^{-\frac32}\log(p) + \Tilde{c}_- p^{-\frac32},\quad p\rightarrow \infty,\label{eq-asymptotic-p0}\end{equation}
for some uniquely defined $\Tilde{c}_+,\Tilde{c}_-$. Then
 {   for $\alpha\neq \frac14$} we set
\begin{align}
   \Tilde{  L}_{\alpha,\mathrm{D}}^{\max}\psi(p)&
    =
    p^2\psi(p)+\Big(\alpha-\frac14\Big)\lim_{\Lambda\to\infty}\Bigg(
    \int_0^\Lambda\mathrm{d}q\big(\theta(p-q)q
    +\theta(q-p)p\big)\psi(q)
+\frac{\Tilde{c}_+p\Lambda^{-\frac12-m}}{\frac12+m}
    +\frac{\Tilde{c}_-p\Lambda^{-\frac12+m}}{\frac12-m}    \Bigg)
    ,\label{mm1o}\\
   \Tilde{  L}_{0,\mathrm{D}}^{\max}\psi(p)&
    =
    p^2\psi(p)-\frac14\lim_{\Lambda\to\infty}\Bigg(
    \int_0^\Lambda\mathrm{d}q\big(\theta(p-q)q
    +\theta(q-p)p\big)\psi(q)
+ \Tilde{c}_+ p \big( 4 \Lambda^{-\frac12} +2\Lambda^{-\frac12}\log\Lambda \big)
    +2 \Tilde{c}_- p\Lambda^{-\frac12}    \Bigg)
    ,\label{mm10o}\\
 \Tilde{  L}_{\alpha,\mathrm{N}}^{\max}\psi(p)&
    =
    p^2\psi(p)-\Big(\alpha-\frac14\Big)\lim_{\Lambda\to\infty}\Bigg(
    \int_0^\Lambda\mathrm{d}q\big(\theta(p-q)p
    +\theta(q-p)q\big)\psi(q)
-\frac{\Tilde{c}_+\Lambda^{\frac12-m}}{\frac12-m}
    -\frac{\Tilde{c}_-\Lambda^{\frac12+m}}{\frac12+m}    \Bigg)
    ,\label{mm2o}\\
   \Tilde{  L}_{0,\mathrm{N}}^{\max}\psi(p)&
    =
    p^2\psi(p){ +}\frac14\lim_{\Lambda\to\infty}\Bigg(
    \int_0^\Lambda\mathrm{d}q\big(\theta(p-q)p
    +\theta(q-p)q\big)\psi(q)
+ \Tilde{c}_+ \big( 4 \Lambda^{\frac12} -2\Lambda^{\frac12}\log\Lambda \big)
    -2 \Tilde{c}_- \Lambda^{\frac12}    \Bigg)
    .\label{mm20o}
\end{align}

Note that in  the Dirichlet case with $\alpha< \frac14$ (and hence
$|\mathrm{Re}(m)|<\frac12$) both counterterms in \eqref{mm1o} are
not necessary---they go to zero and the integrals are
convergent. However in all other cases at least one  counterterm is needed.

  \begin{theorem} {   Let $\alpha<1$ and $\alpha\neq\frac14$.} The operators
$    \Tilde{  L}_{\alpha,\mathrm{D}}^{\max}$ and
  $    \Tilde{  L}_{\alpha,\mathrm{N}}^{\max}$ are well defined closed operators. We have
\begin{align}\label{mnn1a}
  \Tilde{  L}_{\alpha,\mathrm{D}}^{\max}&:= \cF_\mathrm{D} L_\alpha^{\max}\cF_\mathrm{D}
  ,\\\label{mnn2a}
    \Tilde{  L}_{\alpha,\mathrm{N}}^{\max}&:= \cF_\mathrm{N} L_\alpha^{\max} \cF_\mathrm{N}.
\end{align}\label{theo1}
  \end{theorem}

  \noindent{\it Proof.} Consider first the Dirichlet case. It is
  enough to consider
$\psi=\psi_{\pm m,\mathrm{D}}$
          
\begin{align}
&
    p^2\psi(p)+\Big(\alpha-\frac14\Big)\lim_{\Lambda\to\infty}\Bigg(
    \int_0^\Lambda\mathrm{d}q\big(\theta(p-q)q
    +\theta(q-p)p\big)\psi(q)
+\frac{p\Lambda^{-\frac12\mp m}}{\frac12\pm m}
   \Bigg)
         \label{mm1o1}\\
  &\hspace{-10ex}
    =
    p^2\psi(p)+\Big(\alpha-\frac14\Big)\lim_{\Lambda\to\infty}\Bigg(
    \int_p^\Lambda\mathrm{d}q(p-q)\psi(q)
+\frac{p\Lambda^{-\frac12\mp m}}{\frac12\pm
    m}+\frac{\Lambda^{\frac12\mp m}}{\frac12\mp m}
      \Bigg)
            \label{mm1o2}\\
  &
    =
    p^2\big(\psi(p)-p^{-\frac32\mp m}\big)
    +\Big(\alpha-\frac14\Big)
    \int_p^\infty\mathrm{d}q(p-q)\big(\psi(q)-q^{-\frac32\mp m}\big)
            .\label{mm1o3}
\end{align}
Thus $   \Tilde{  L}_{\alpha,\mathrm{D}}^{\max}\psi(p)$ is well defined.

Using Lemma \ref{ppen} 1 and \eqref{mm1o2}
 we show that for $x>0$, possibly in  the sense of an
oscillatory integral,
\begin{align}
  &\sqrt{\frac2\pi}\int_0^\infty\sin(px)\big(\Tilde{L}_{\alpha,\mathrm{D}}^{\max}\psi\big)(p)
    \mathrm{d}p
  \\
    =&\Bigg(-\partial_x^2+\Big(\alpha-\frac14\Big) \frac1{x^2}\Bigg)  
     \sqrt{\frac2\pi}\int_0^\infty\sin(px) 
     \psi(p) \mathrm{d}p,
  \end{align}
  which proves \eqref{mnn1a}.
  
The Neumann case is analogous. It is
  enough to consider
$\psi=\psi_{\pm m,\mathrm{N}}$.
          
\begin{align}
&
    p^2\psi(p)-\Big(\alpha-\frac14\Big)\lim_{\Lambda\to\infty}\Bigg(
    \int_0^\Lambda\mathrm{d}q\big(\theta(p-q)p
    +\theta(q-p)q\big)\psi(q)
-\frac{\Lambda^{\frac12\mp m}}{\frac12\mp  m}
   \Bigg)
         \label{mm1o21}\\
  &\hspace{-10ex}
    =
    p^2\psi(p)+\Big(\alpha-\frac14\Big)\lim_{\Lambda\to\infty}\Bigg(
    \int_p^\Lambda\mathrm{d}q(p-q)\psi(q)
+\frac{p\Lambda^{-\frac12\mp m}}{\frac12\pm
    m}+\frac{\Lambda^{\frac12\mp m}}{\frac12\mp m}
      \Bigg)
            \label{mm1o22}\\
  &
    =
    p^2\big(\psi(p)-p^{-\frac32\mp m}\big)
    +\Big(\alpha-\frac14\Big)
    \int_p^\infty\mathrm{d}q(p-q)\big(\psi(q)-q^{-\frac32\mp m}\big)
            .\label{mm1o23}
\end{align}
Thus $   \Tilde{  L}_{\alpha,\mathrm{N}}^{\max}\psi(p)$ is well defined.

Using Lemma \ref{ppen} 2 and \eqref{mm1o22}
 we show that for $x>0$,
\begin{align}
  &\sqrt{\frac2\pi}\int_0^\infty\cos(px)\big(\Tilde{L}_{\alpha,\mathrm{N}}^{\max}\psi\big)(p)
    \mathrm{d}p
  \\
    =&\Bigg(-\partial_x^2+\Big(\alpha-\frac14\Big) \frac1{x^2}\Bigg)  
     \sqrt{\frac2\pi}\int_0^\infty\cos(px) 
     \psi(p) \mathrm{d}p,
  \end{align}
  which proves \eqref{mnn2a}. The special case $m=0$ is proven in an analogous way. \qed

Let us remark  that the ``kinetic terms'' for $ p^{-\frac32\pm m}$, that is, $p^2 p^{-\frac32\pm m}$, are never square integrable for $|m|<1$. Therefore, to obtain an element of $L^2$ one needs to balance them with the integral terms.

{   Consider now $\alpha=\frac14$ and $m=\frac12$.
In the position space this corresponds to  the free
Schr\"odinger operator on the half-line and the description of the
maximal operator is straightforward.
In the momentum space the situation is
more problematic.  The
previous construction does not work.
%Below we describe a construction of $ \Tilde{L}_{\frac14,\mathrm{D}}^{\max}$
%and $ \Tilde{L}_{\frac14,\mathrm{N}}^{\max}$ in the  momentum approach.

Introduce the space
\begin{equation}\label{amm11}
 L^{2,2}(\rr_+)  :=\left\{\psi\in L^{2}(\rr_+)\ |\
\int_0^\infty|\psi(p)|^2p^4\d p<\infty\right\}.
\end{equation}

Let us choose $  \psi_\mathrm{D},  \psi_\mathrm{N}\in
L^{2}(\mathbb{R}_+)$
satisfying the following condition: there exists $\Lambda_0>0$ such
that
\begin{align}\label{pqw1o}
  p>\Lambda_0\ \Rightarrow\quad
  \psi_{\mathrm{D}}(p)=p^{-1},\\
  \label{pqw2o}
  p>\Lambda_0\ \Rightarrow\quad
  \psi_{\mathrm{N}}(p)=p^{-2}.\end{align}
We set
\begin{align}
  \cD
  (  \Tilde{L}_{\frac14,\mathrm{D}}^{\max})&:= L^{2,2}(\rr_+) +
                                            \mathbb{C}\psi_{\mathrm{D}};\label{wna1o}\\
   \cD
  (
  \Tilde{L}_{\frac14,\mathrm{N}}^{\max})&:= L^{2,2}(\rr_+) +
                                                        \mathbb{C}\psi_{\mathrm{N}}.
\label{wna2o}
\end{align}
Thus every $\psi$ in $ \cD  (  \Tilde{L}_{\frac14,\mathrm{D}}^{\max})$
or in $ \cD  (  \Tilde{L}_{\frac14,\mathrm{N}}^{\max})$ behaves like
\begin{align}\psi(p) &\sim \Tilde{c} p^{-1}, \quad  p\rightarrow \infty;
                       \label{eq-asymptotic-po}\\
  \text{resp.}\qquad \psi(p) &\sim \Tilde{c}
                                      p^{-2}, \quad  p\rightarrow \infty,\label{eq-asymptotic-p0o}\end{align} 
for some uniquely defined $\Tilde{c}$. Then
we set
\begin{align}
   \Tilde{  L}_{\frac14,\mathrm{D}}^{\max}\psi(p)&
    =
    p^2\big(\psi(p)-\Tilde cp^{-1}\big),\label{mm1oo}\\
\text{resp.}\qquad \Tilde{  L}_{\frac14,\mathrm{N}}^{\max}\psi(p)&
    =                                        p^2\big(\psi(p)-\Tilde cp^{-2}\big)
                          .\label{mm2oo}
\end{align}

Now we can extend Theorem \ref{theo1} to $\alpha=\frac14$:                                  
\begin{theorem}\label{theo3}
The operators $\Tilde   L_{\frac14,\mathrm{D}}^{\max}$ and 
$\Tilde   L_{\frac14,\mathrm{D}}^{\max}$ are well defined and
closed. We have                     
\begin{align}\label{mnn1a}
  \Tilde{  L}_{\frac14,\mathrm{D}}^{\max}&:= \cF_\mathrm{D} L_\frac14^{\max}\cF_\mathrm{D}
  ,\\\label{mnn2a}
    \Tilde{  L}_{\frac14,\mathrm{N}}^{\max}&:= \cF_\mathrm{N} L_\frac14^{\max} \cF_\mathrm{N}.
\end{align}
\end{theorem}

\textit{Proof.}
First we note that
\begin{equation}
  \Tilde   L_{\frac14,\mathrm{D}}^{\max}\Big|_{L^{2,2}(\mathbb{R}_+)}=
  \Tilde
  L_{\frac14,\mathrm{N}}^{\max}\Big|_{L^{2,2}(\mathbb{R}_+)}\end{equation}
coincides with the multiplication by $p^2$, and also with
$\cF_\mathrm{D} H_\mathrm{D}\cF_\mathrm{D}$ and
$\cF_\mathrm{N} H_\mathrm{N}\cF_\mathrm{N}$.

Let $\epsilon>0$.
We set
\begin{equation}\label{phi}
  \phi_\epsilon(x)=\e^{-\epsilon x},\end{equation} which belongs to
$\cD(L_\frac14^{\max})$ but does not belong to
$\cD(H_\mathrm{D})$ or $\cD(H_\mathrm{N})$.
We have 
\begin{align}\sqrt{\frac\pi2}\cF_\mathrm{D}\phi_\epsilon(p)=\frac{p}{p^2+\epsilon^2}=:\psi_{\mathrm{D},\epsilon}(p) \quad\in\quad 
  p^{-1}+
L^{2,2}(\mathbb{R}_+);\\
\frac{1}{\epsilon}\sqrt{\frac\pi{2}}  \cF_\mathrm{N}\phi_\epsilon(p)=\frac{1}{p^2+\epsilon^2}=:\psi_{\mathrm{N},\epsilon}(p)
  \quad\in\quad 
  p^{-2}+
L^{2,2}(\mathbb{R}_+).
  \end{align}
Clearly, $L_{\frac14}^{\max}\phi_\epsilon=-\epsilon^2\phi_\epsilon$
and hence
\begin{align}
\cF_\mathrm{D}
  L_\frac14^{\max}\cF_\mathrm{D}\psi_{\mathrm{D},\epsilon}(p)&
                                                               =-\epsilon^2 \psi_{\mathrm{D},\epsilon}(p)=-p+p^2 \psi_{\mathrm{D},\epsilon}(p);\\
\cF_\mathrm{N}
  L_\frac14^{\max}\cF_\mathrm{N}\psi_{\mathrm{N},\epsilon}(p)&
                                                               =-\epsilon^2 \psi_{\mathrm{N},\epsilon}(p)=-1+p^2 \psi_{\mathrm{N},\epsilon}(p).   
  \end{align}
Therefore,  $\psi_{\mathrm{D},\epsilon}(p)$ and  $\psi_{\mathrm{N},\epsilon}(p)$ satisfy
\eqref{mm1oo}, resp. \eqref{mm2oo}. $\Box$

}
                                 
  \subsection{Self-adjoint realizations in the momentum approach 
       I}

  {   As announced in the introduction, we will now
    describe two constructions of self-adjoint realizations of Bessel
    operators in the momentum approach. 
  In this subsection we describe the first, which we called
  ``mathematicians' approach''. It involves  
  selecting an appropriate domain inside the maximal domain.
  % We will   restrict ourselves to the case $\alpha\neq\frac14$, which 
%allows us to use the description of the maximal operator given by 
  %\eqref{wna1}--\eqref{wna20}. 

  We will first discuss the case  $\alpha\neq\frac14$.
 Assume that $\alpha<1$ and $m^2=\alpha$. Let $\tilde\xi\in
 C^\infty(\mathbb{R}_+)$
 such that $\tilde\xi=1$ near $\infty$. Let
$\Tilde\kappa,\Tilde\nu\in\mathbb{C}$. Clearly, the following
subspaces do not
depend on the choice of $\Tilde\xi$:}
  \begin{equation*}
\begin{aligned}
&\mathcal{D}(\Tilde{H}_{m,\Tilde{\kappa},\mathrm{D/N}}):= \{ g \in
\mathcal{D}(\Tilde{L}^{\mathrm{max}}_{m^2,\mathrm{D/N}}) \quad | \quad
\text{for some} \:  c \in \mathbb{C} \quad g(p) - c\tilde\xi(p) ( p^{-3/2 -m} + \Tilde{\kappa} p^{-3/2 + m}) \in \mathcal{D}(\Tilde{L}^{\mathrm{min}}_{m^2,\mathrm{D/N}})  \},\\
&\mathcal{D}(\Tilde{H}_{m,\infty,\mathrm{D/N}}) : = \{ g \in
\mathcal{D}(\Tilde{L}^{\mathrm{max}}_{m^2,\mathrm{D/N}}) \quad | \quad
\text{for some} \:  c \in \mathbb{C} \quad g(p) - c\tilde\xi(p) p^{-3/2 +m} \in \mathcal{D}(\Tilde{L}^{\mathrm{min}}_{m^2,\mathrm{D/N}})  \},\\
&\mathcal{D}(\Tilde{H}_{0,\mathrm{D/N}}^{\Tilde{\nu}}):= \{ g \in
\mathcal{D}(\Tilde{L}^{\mathrm{max}}_{0,\mathrm{D/N}}) \quad | \quad
\text{for some} \:  c \in \mathbb{C} \quad g(p) - c\tilde\xi(p)(p^{-3/2}\ln p + \Tilde{\nu} p^{-3/2}) \in \mathcal{D}(\Tilde{L}^{\mathrm{min}}_{0,\mathrm{D/N}})\} ,\\
&\mathcal{D}(\Tilde{H}_{0,\mathrm{D/N}}^{\infty}) := \{ g \in
\mathcal{D}(\Tilde{L}^{\mathrm{max}}_{0,\mathrm{D/N}}) \quad | \quad
\text{for some} \: c \in \mathbb{C} \quad g(p) - c \tilde\xi(p)p^{-3/2} \in \mathcal{D}(\Tilde{L}^{\mathrm{min}}_{0,\mathrm{D/N}})  \}.
\end{aligned}
\end{equation*}

  We  define the operators $\Tilde{H}_{m,\Tilde{\kappa},\mathrm{D}}$,
    $\Tilde{H}_{m,\Tilde{\kappa},\mathrm{N}}$,
       $\Tilde{H}_{m,\mathrm{D}}^{\Tilde{\nu}}$,        $\Tilde{H}_{m,\mathrm{N}}^{\Tilde{\nu}}$,
  to be the operators $\Tilde{L}_{m^2,\mathrm{D}}^{\max}$, resp.
   $\Tilde{L}_{m^2,\mathrm{N}}^{\max}$ restricted to the domains described above.

  \begin{theorem}\label{Th-3} {   Let  $m^2<1$ and $m\neq\pm\frac12$.} The above defined operators
    are self-adjoint in the following situations:
\begin{itemize}
\item $\Tilde{H}_{m,\Tilde\kappa,\mathrm{D}}$,
   $\Tilde{H}_{m,\Tilde\kappa,\mathrm{N}}$
  for $m\in ]-1,1[\backslash\{0\}$ and $\Tilde{\kappa} \in \mathbb{R}\cup\{\infty\}$,
    \item  $\Tilde{H}_{m,\Tilde\kappa,\mathrm{D}}$,
   $\Tilde{H}_{m,\Tilde\kappa,\mathrm{N}}$ for  $m \in \mathrm{i}\mathbb{R}\backslash\{0\}$ and $|\Tilde\kappa|=1$,
    \item $\Tilde{H}_{0,\mathrm{D}}^{\Tilde{\nu}}$,  $\Tilde{H}_{0,\mathrm{N}}^{\Tilde{\nu}}$
      for  $\nu \in \mathbb{R}\cup\{\infty\}$.
\end{itemize}
Besides, {   denoting by $\gamma$  the Euler constant, we have}
  \begin{align}
    \cF_\mathrm{D}H_{m,\kappa}    \cF_\mathrm{D}&=\Tilde{H}_{m, \Tilde{\kappa}_\mathrm{D},\mathrm{D}},\qquad
    \Tilde{\kappa}_\mathrm{D}=\frac{\sin(\frac\pi4+\frac{\pi m}2)\Gamma(-\frac12-m)}{\sin(\frac\pi4-\frac{\pi m}2)\Gamma(-\frac12+m)}\kappa,\\
        \cF_\mathrm{N}H_{m,\kappa}    \cF_\mathrm{N}&=\Tilde{H}_{m,\Tilde{\kappa}_\mathrm{N}, \mathrm{N}},\qquad
    \Tilde{\kappa}_\mathrm{N}=\frac{\cos(\frac\pi4+\frac{\pi m}2)\Gamma(-\frac12-m)}{\cos(\frac\pi4-\frac{\pi m}2)\Gamma(-\frac12+m)}\kappa,\\
    \cF_\mathrm{D}H_0^\nu    \cF_\mathrm{D}&=\Tilde{H}_{0,\mathrm{D}}^{\Tilde{\nu}_\mathrm{D}},\qquad
    \Tilde{\nu}_\mathrm{D}= -\nu - 2 + \gamma + \frac{\pi}{2} + \log 4,\\
    \cF_\mathrm{N}H_0^\nu    \cF_\mathrm{N}&=\Tilde{H}_{0,\mathrm{N}}^{\Tilde{\nu}_\mathrm{N}},\qquad
    \Tilde{\nu}_\mathrm{N}= -\nu - 2 + \gamma - \frac{\pi}{2} + \log 4.
  \end{align}\label{theo}
\end{theorem}
%{  In the degenerate cases $m=\pm\frac12$ one has: for $m=+\frac12$ $\Tilde{\kappa}_{\mathrm{D}}=\infty$, $\Tilde{\kappa}_{\mathrm{N}}=0$, and for $m=-\frac12$  $\Tilde{\kappa}_{\mathrm{D}}=0$, $\Tilde{\kappa}_{\mathrm{N}}=\infty$.}\\
{  
\textit{Proof.}}
The equalities above follow from sine and cosine transforms \eqref{eq-sin-tr} and \eqref{eq-cos-tr}, and the definition \eqref{eq-domains}. The limiting case $m=0$ is obtained by considering the limit $m\rightarrow 0$ with $\kappa = -1+2m\nu$ and $\Tilde{\kappa}_{\mathrm{D/N}} = -1-2m\Tilde{\nu}_{\mathrm{D/N}}$.\footnote{Note that for $m\rightarrow 0$ one has $(-1+2m\nu)x^{-m}+x^m \sim -2m(\nu+\log x)$. } {   Integration by parts shows that on considered domains the discussed operators are indeed self-adjoint. $\Box$.

%In the introduction the above theorem was called
  %``mathematicians' style construction of Bessel operators in the momentum
  %representation''.
%  It can be proven in many ways--by a reduction to the
  % position representation, or directly in the momentum
   %representation. 

The case $\alpha=\frac14$ needs a separate treatment. Let
$\kappa\in\mathbb{R}\cup\{0\}$, $\kappa\neq0$. Choose functions
$\psi_{\mathrm{D},\underline\kappa},
\psi_{\mathrm{D},\underline\kappa}\in L^2(\mathbb{R}_+)$ satisfying
the following conditions:
\begin{align}\label{pqw1ou}
  p>\Lambda_0\ \Rightarrow\quad
 & \psi_{\mathrm{D},\underline\kappa}(p)=p^{-1},\qquad
\int_0^\infty p\big(
  \psi_{\mathrm{D},\underline\kappa}(p)-p^{-1}\big)\d p=\frac{\pi}{2\kappa};
  \\
  \label{pqw2ou}
  p>\Lambda_0\ \Rightarrow\quad
  &\psi_{\mathrm{N},\underline \kappa}(p)=p^{-2},\qquad
\int_0^\infty 
  \psi_{\mathrm{N},\underline\kappa}(p)\d p=-\frac{\pi}{2\kappa}.\end{align}
(The underline below $\kappa$ is meant to stress a different role of
$\kappa$ in the case $\alpha=\frac14$ than that of $\Tilde\kappa$ for
$\alpha\neq\frac14$). We set
\begin{align}
  \cD(\Tilde{H}_{\frac12,\underline0,\mathrm{D}})&:=L^{2,2}(\mathbb{R}_+);\\
    \cD(\Tilde{H}_{\frac12,\underline\kappa,\mathrm{D}})&:=\cD(\Tilde{L}_{\frac14,\mathrm{D}}^{\min})+\mathbb{C}
                                                          \psi_{\mathrm{D},\underline\kappa},\quad\kappa\neq0;\\
  \cD(\Tilde{H}_{-\frac12,\underline\kappa,\mathrm{D}})&:=
                                                           \cD(\Tilde{H}_{\pm\frac12,\underline\kappa^{-1},\mathrm{D}})
      ;\\    \cD(\Tilde{H}_{-\frac12,\underline0,\mathrm{N}})&:=L^{2,2}(\mathbb{R}_+);\\
    \cD(\Tilde{H}_{-\frac12,\underline\kappa,\mathrm{N}})&:=\cD(\Tilde{L}_{\frac14,\mathrm{N}}^{\min})+\mathbb{C}
                                                         \psi_{\mathrm{N},\underline\kappa},\quad\kappa\neq0;\\
\cD(\Tilde{H}_{\frac12,\underline\kappa,\mathrm{N}})&:=
                                                          \cD(\Tilde{H}_{-\frac12,\underline\kappa^{-1},\mathrm{N}}).
\end{align}
We define $\Tilde{H}_{\pm\frac12,\underline\kappa,\mathrm{D}}$
  and  $\Tilde{H}_{\pm\frac12,\underline\kappa,\mathrm{N}}$ to be the
    restrictions of 
 $\Tilde{L}_{\pm\frac14,\mathrm{D}}$
  resp.  $\Tilde{L}_{\pm\frac14,\mathrm{N}}$ to the corresponding
    domains specified above.

    \begin{theorem}
      For $\kappa\in\mathbb{R}\cup\{\infty\}$ the above operators are
      self-adjoint and
       \begin{align}
    \cF_\mathrm{D}H_{\pm\frac12,\kappa}    \cF_\mathrm{D}&=\Tilde{H}_{\pm\frac12,\underline\kappa, \mathrm{D}},\\
        \cF_\mathrm{N}H_{\pm\frac12,\kappa}    \cF_\mathrm{N}&=\Tilde{H}_{\pm\frac12,\underline\kappa,\mathrm{N}}.
  \end{align}\label{theo0}
\end{theorem}

\textit{Proof.} {  
It is enough to show
       \begin{align}
    \cF_\mathrm{D}H_{\frac12,\kappa}    \cF_\mathrm{D}&=\Tilde{H}_{\frac12,\underline\kappa, \mathrm{D}},\\
        \cF_\mathrm{N}H_{-\frac12,\kappa}    \cF_\mathrm{N}&=\Tilde{H}_{-\frac12,\underline\kappa,\mathrm{N}}
  \end{align}
for $\kappa\neq0$.

Let us first consider the case $\kappa < 0$, which is simpler. Every function in $ \mathcal{D}(H_{\frac12, \kappa} ) = \mathcal{D}( H_{-\frac12, 1/\kappa} )$  can be unambiguously written as a sum of an element of the minimal domain and $c e^{-\epsilon x}$, where $\epsilon = -1/\kappa>0$ and $c\in \mathbb{C}$.
% Recall from   \eqref{phi}
%Let $\epsilon>0$. The vector $
% \phi_\epsilon(x)=
%\e^{-\epsilon x}$ belongs to $\cD(H_{\frac12,\kappa})$ with
% $\kappa=-\frac1\epsilon$.
The  sine transform of $ \e^{-\epsilon x}$ is proportional to
    $\frac{p}{p^2+\epsilon^2}$. Now
    \begin{equation}
      \int_0^\infty\Big(p \frac{p}{p^2+\epsilon^2}-1\Big)\d
      p=-\epsilon\frac\pi2=\frac\pi{2\kappa}.
    \end{equation}
    Hence $ \frac{p}{p^2+\epsilon^2}$ belongs
    to $\cD(\Tilde{H}_{\frac12,\underline\kappa, \mathrm{D}})$.

    % $\phi_\epsilon$
%belongs also to $\cD(H_{-\frac12,\kappa})$ with
%$\kappa=-\epsilon$.
    The cosine transform of     $\e^{-\epsilon x}$ is proportional to
    $\frac{1}{p^2+\epsilon^2}$, and
    \begin{equation}
      \int\frac{1}{p^2+\epsilon^2}\d
      p=\frac\pi{2\epsilon}=-\frac\pi{2\kappa}.
    \end{equation}
    Hence $ \frac{1}{p^2+\epsilon^2}$ belongs
    to $\cD(\Tilde{H}_{-\frac12,\underline\kappa, \mathrm{N}})$.

To obtain the analogous decomposition of vectors from  $
\mathcal{D}(H_{\frac12, \kappa} ) = \mathcal{D}( H_{-\frac12,
  1/\kappa} )$ for $\kappa > 0$ and $\kappa = \infty$, we consider
a function $2 \e^{-x} - \e^{-\delta x}$.
 for $\delta \geq 2$.
Near $x=0$ it behaves like $1 + (\delta - 2) x$, so that it is an element of $ \mathcal{D}(H_{\frac12, \kappa} ) = \mathcal{D}( H_{-\frac12, 1/\kappa} )$ for $\kappa = \frac{1}{\delta - 2}$.
Its sine transform is proportional to $2\frac{p}{p^2 + 1} - \frac{p}{p^2+\delta^2}$, and
\begin{equation}
\int_0^\infty \Big( 2\frac{p}{p^2 + 1} - \frac{p}{p^2+\delta^2} - 1 \Big) = \frac{\pi}{2} (\delta - 2 ) = \frac{\pi}{2\kappa}.
\end{equation}
Its cosine transform is proportional to $2\frac{1}{p^2 + 1} - \frac{1}{p^2+\delta^2}$, and
\begin{equation}
\int_0^\infty \Big( 2\frac{1}{p^2 + 1} - \frac{1}{p^2+\delta^2}  \Big) = \frac{\pi}{2} (2 - \delta ) = -\frac{\pi}{2\kappa}.
\end{equation}

The case $\kappa = 0$ is treated in the same way, by taking e.g.  $\e^{-x} - \e^{-2x}$.
%Since any vector in $ \mathcal{D}(H_{\frac12, \kappa} ) = \mathcal{D}( H_{-\frac12, 1/\kappa} )$ can be decomposed as a sum of considered exponents and functions from the minimal domain, and, conversely, any vector from $\cD(\Tilde{H}_{-\frac12,\underline\kappa, \mathrm{D/N}})$ can be written as a sum of sine/cosine transforms of those functions and elements of $\cD(\Tilde{L}_{\frac14,\mathrm{D/N}}^{\min})$, one obtains
%\begin{align}
%& \cF_\mathrm{D}H_{\pm\frac12,\kappa}    \cF_\mathrm{D} = \Tilde{H}_{\pm\frac12,\underline\kappa, \mathrm{D}}, \\
%& \cF_\mathrm{N}H_{\pm\frac12,\kappa}    \cF_\mathrm{N} = \Tilde{H}_{\pm\frac12,\underline\kappa, \mathrm{N}}.
%\end{align}
%Theorem \ref{theo3} ensures, that the discussed operators are equal.
$\Box$
}}

\subsection{    Self-adjoint realizations in the momentum approach 
   II}

  We will now present what we called ``physicists' style
  construction'' of selfa-adjoint Bessel  operators. As we discussed in the introduction, this construction
can be viewed as a {  ``toy
  illustration'' of the Wilsonian renormalization group.
It will be valid for all $\alpha<1$, including $\alpha=\frac14$.}

The first step of ``physicists' construction'' is imposing a cut  off
on
\eqref{mm1} and \eqref{mm2}:
\begin{align}\label{mmq1}
	&\Tilde{L}_{\alpha,\mathrm{D}} (\Lambda) \psi(p) := \theta(\Lambda-p) \bigg( p^2\psi(p)+\Big(\alpha-\frac14\Big)\int_0^\Lambda\big(\theta(p-q)q
    +\theta(q-p)p\big)\psi(q) \mathrm{d}q \bigg), \\\label{mmq2}
	&\Tilde{L}_{\alpha,\mathrm{N}} (\Lambda) \psi(p) := \theta(\Lambda-p) \bigg( p^2\psi(p)-\Big(\alpha-\frac14\Big)\int_0^\Lambda\big(\theta(p-q)p
    +\theta(q-p)q\big)\psi(q) \mathrm{d}q \bigg).
\end{align}
Note that \eqref{mmq1} and \eqref{mmq2} are bounded and self-adjoint.
We also need
 counterterms, which  will be proportional to the following rank-one operators:
\begin{align}
  K_\mathrm{D}(\Lambda)\psi(p)&:=\theta(\Lambda-p)\int_0^\Lambda pq
  \psi(q)\d q,\\
    K_\mathrm{N}(\Lambda)\psi(p)&:=\theta(\Lambda-p)\int_0^\Lambda 
    \psi(q)\d q.
    \end{align}
To gain the intuition about the above operators let us note that
for a large class of functions $\psi$,
\begin{align}
  \lim_{\Lambda\to\infty}(\cF_\mathrm{D}\psi|K_\mathrm{D}(\Lambda)\cF_\mathrm{D}\psi)
  &  =\frac{1}{2\pi}|\psi'(0)|^2,\\
   \lim_{\Lambda\to\infty}(\cF_\mathrm{N}\psi|K_\mathrm{N}(\Lambda)\cF_\mathrm{N}\psi)
&  =\frac{1}{2\pi}|\psi(0)|^2.
\end{align}

We have already noted an important role played by the dilation in the
position representation. In principle, we have two kinds of momentum
representations--obtained by the sine and cosine transformation. The dilation coincides for both kinds of the momentum representation, and it will be denoted by
 $\Tilde{U}(\tau)$:
\begin{equation}
  \Tilde{U}(\tau)
  =\cF_\mathrm{D}U(\tau)\cF_\mathrm{D}^{-1}=\cF_\mathrm{N}U(\tau)\cF_\mathrm{N}^{-1}.
  \end{equation}
Note that it acts in the opposite way than in the position representation:
\begin{equation}
\Tilde{U}(\tau)\psi(p)=\e^{-\frac{1}{2}\tau}\psi(\e^{-\tau}p).
  \end{equation}
It is easy to see that
\begin{align}
\Tilde{U}(\tau)\Tilde{L}_{\alpha,\mathrm{D}}(\Lambda)\Tilde{U}(-\tau)
  =\e^{-2\tau}\Tilde{L}_{\alpha,\mathrm{D}}(\e^\tau\Lambda),&\qquad
     \Tilde{U}(\tau)K_\mathrm{D}(\Lambda)\Tilde{U}(-\tau)
  =\e^{-3\tau}K_\mathrm{D}(\e^\tau\Lambda);\label{pqu1}\\
  \Tilde{U}(\tau)\Tilde{L}_{\alpha,\mathrm{N}}(\Lambda)\Tilde{U}(-\tau)
  =\e^{-2\tau}\Tilde{L}_{\alpha,\mathrm{N}}(\e^\tau\Lambda),&\qquad
   \Tilde{U}(\tau)K_\mathrm{N}(\Lambda)\Tilde{U}(-\tau)
  =\e^{-\tau} K_\mathrm{N}(\e^\tau\Lambda)  .\label{pqu2}
\end{align}

We still need to multiply $K_\mathrm{D}$ and 
$K_\mathrm{N}$ by cutoff-dependent (``running'') coupling constants.
%This suggests to multiply $K_\mathrm{D}$ by $\frac{1}{\Lambda}$ and 
%$K_\mathrm{N}$ by $\Lambda$.
{   \eqref{pqu1} and \eqref{pqu2} suggest to write} these  coupling
constants as
$\frac{f_\Lambda}{\Lambda}$, resp.
$\Lambda g_\Lambda$. Thus
we expect that for suitably chosen functions $\Lambda\mapsto f_\Lambda,g_\Lambda$, the following operators approximate
 our Hamiltonians:
  \begin{align}
&    \Tilde{  L}_{\alpha,\mathrm{D}}(\Lambda)+\frac{f_\Lambda}{\Lambda}
K_\mathrm{D}(\Lambda)
  ,\label{mma1b}\\
 &   \Tilde{  L}_{\alpha,\mathrm{N}}(\Lambda)
+\Lambda g_\Lambda
K_\mathrm{N}(\Lambda)
.\label{mma2b}
  \end{align}

Let us guess  the dependence
of $f_\Lambda$ and $g_\Lambda$  on the cut-off $\Lambda$. 
Assume that for large momenta $\psi(p)\sim p^{-\frac32-m}+\Tilde{\kappa}p^{-\frac32+m}$.
Acting with \eqref{mma1b} and \eqref{mma2b} on $\psi$
for large $p<\Lambda$ we obtain 
the following leading-order dependence on the cut-off:\begin{align}
   &    \Big(\Tilde{  L}_{\alpha,\mathrm{D}}(\Lambda)+\frac{f_\Lambda}{\Lambda}
K_\mathrm{D}(\Lambda)\Big)\psi(p)
,\label{mma1c}\\
\sim&p\Big(m^2-\frac14\Big)\Big(\frac{\Lambda^{-\frac12-m}}{(-\frac12-m)}
    +\Tilde\kappa\frac{\Lambda^{-\frac12+m}}{(-\frac12+m)}\Big)
+p f_\Lambda\Big(\frac{\Lambda^{-\frac12-m}}{(\frac12-m)}
+\Tilde\kappa\frac{\Lambda^{-\frac12+m}}{(\frac12+m)}\Big)\label{jk1}
,\\
 &  \big( \Tilde{  L}_{\alpha,\mathrm{N}}(\Lambda)
+\Lambda g_\Lambda
K_\mathrm{N}(\Lambda)\big)\psi(p)
\label{mma2c}\\
 \sim &-\Big(m^2-\frac14\Big)\Big(\frac{\Lambda^{\frac12-m}}{(\frac12-m)}
    +\Tilde\kappa\frac{\Lambda^{\frac12+m}}{(\frac12+m)}\Big)
+g_\Lambda\Big(\frac{\Lambda^{\frac12-m}}{(\frac12+m)}
    +\Tilde\kappa\frac{\Lambda^{\frac12+m}}{(\frac12-m)}\Big)  .\label{jk2}
    \end{align}

Demanding that \eqref{jk1} and \eqref{jk2} vanish yields
\begin{align}
  f_{m,\Tilde{\kappa},\Lambda}&=-\frac{\Big((\frac12-m)\Lambda^{-m}
    +\Tilde\kappa(\frac12+m)\Lambda^{m}\Big)}
{\Big((\frac12-m)^{-1}\Lambda^{-m}
  +\Tilde\kappa(\frac12+m)^{-1} \Lambda^{m}\Big)},\label{eq-f-Lambda}\\ 
&=-\frac14-m^2-m\tanh\big(m\log(\Lambda/\lambda_\mathrm{D})\big),\qquad \lambda_\mathrm{D}^{-2m}=\frac{(\frac12-m)}{(\frac12+m)}\tilde\kappa;\label{eq-f-Lambda}\\
g_{m,\Tilde{\kappa},\Lambda}&={   -}\frac{\Big((\frac12+m)\Lambda^{-m}
  +\Tilde\kappa(\frac12-m)\Lambda^{m}\Big)}
{\Big((\frac12+m)^{-1}\Lambda^{-m}
  +\Tilde\kappa(\frac12-m)^{-1} \Lambda^{m}\Big)},\label{eq-g-Lambda}\\ 
&={   -}\frac14{   -}m^2{   +}m\tanh\big(m\log(\Lambda/\lambda_\mathrm{N})\big),\qquad \lambda_\mathrm{N}^{{   -}2m}=\frac{(\frac12+m)}{(\frac12-m)}\tilde\kappa. \label{eq-g-Lambda}
\end{align}
{   It is easy to check that $ g_{m,\Tilde{\kappa},\Lambda}$ can be obtained from $f_{m,\Tilde{\kappa},\Lambda}$ by changing $m\rightarrow -m$ and $\Lambda\rightarrow \Lambda^{-1}$.} In terms of an initial value at the scale $\Lambda_0$ we can write
\begin{align}\label{eqq1}
  f_{m,\Tilde{\kappa},\Lambda}
  &=-\frac14-m^2-m\frac{\big((\frac12-m)^2+f_{\Lambda_0}\big)\Lambda^{2m}
      +\big((\frac12+m)^2+f_{\Lambda_0}\big)\Lambda_0^{2m}}
      {-\big((\frac12-m)^2+f_{\Lambda_0}\big)\Lambda^{2m}
+\big((\frac12+m)^2+f_{\Lambda_0}\big)\Lambda_0^{2m}};\\\label{eqq2}
      g_{m,\Tilde{\kappa},\Lambda}&=
{   -}\frac14{   -}m^2+m\frac{\big((\frac12+m)^2{   +}g_{\Lambda_0}\big)\Lambda^{2m}
      +\big((\frac12-m)^2{   +}g_{\Lambda_0}\big)\Lambda_0^{2m}}
      {{   -}\big((\frac12+m)^2{   +}g_{\Lambda_0}\big)\Lambda^{2m}
+\big((\frac12-m)^2{   +}g_{\Lambda_0}\big)\Lambda_0^{2m}}.
\end{align}

For $m=0$ the parameter $\Tilde{\kappa}$ does not work. Applying the de l'Hopital rule with  $\Tilde{\kappa}=-1-2m\Tilde\nu$ we obtain
\begin{align} f_{0,\Lambda}^{\Tilde{\nu}}&=-   \frac{(\log\Lambda+\Tilde{\nu}+2)}{4(\log\Lambda+\Tilde{\nu}-2)}
 \label{eqq1a} ,\\ g_{0,\Lambda}^{\Tilde{\nu}}&=  -\frac{(\log\Lambda+\Tilde{\nu}-2)}{4(\log\Lambda+\Tilde{\nu}+2)}.\label{eqq2a}
\end{align}

Note that (for both $m\neq0$ and $m=0$)   $f_\Lambda$ and $g_\Lambda$ satisfy the differential equations 
\begin{align}\label{eq1}
  \Lambda\frac{\mathrm{d}}{\mathrm{d}\Lambda} f_\Lambda 
  &=\Big(f_\Lambda+\frac14+m^2\Big)^2-m^2;\\\label{eq2}
    \Lambda\frac{\mathrm{d}}{\mathrm{d}\Lambda} g_\Lambda 
  &={   -}\Big(g_\Lambda{  +}\frac14{   +}m^2\Big)^2{  +}m^2. 
\end{align}

The above heuristic analysis  can be transformed into  the following
rigorous statement:
\begin{theorem}\label{theorem-final}
  Let $m^2<1$. Let $f_\Lambda$ and $g_\Lambda$ be solutions of equations
  \eqref{eq1}, resp. \eqref{eq2}. Then the  bounded self-adjoint operators
    \begin{align}
&    \Tilde{  L}_{m^2,\mathrm{D}}(\Lambda)+\frac{f_\Lambda}{\Lambda}
K_\mathrm{D}(\Lambda)
  ,\label{mma1d}\\
 &   \Tilde{  L}_{m^2,\mathrm{N}}(\Lambda)
+\Lambda g_{\Lambda}
K_\mathrm{N}(\Lambda)
\label{mma2d}
  \end{align}
  converge as $\Lambda\to\infty$ to one of self-adjoint realizations of
  Bessel operators $\Tilde{L}_{m^2}$   in the strong resolvent sense.

  In
  particular, for $m^2\neq\frac14$  take 
  $f_\Lambda=f_{m,\Tilde{\kappa},\mathrm{D}}$ and 
  $g_\Lambda=g_{m,\Tilde{\kappa},\mathrm{N}}$ as in
  \eqref{eqq1}, \eqref{eqq1a}, resp.  \eqref{eqq2}, \eqref{eqq2a}. Then these limits are 
    $\Tilde{H}_{m,\Tilde{\kappa},\mathrm{D}}$, resp.    $\Tilde{H}_{m,\Tilde{\kappa},\mathrm{N}}$.
%    {  Let $m$ and $\Tilde{\kappa}$ be as in Theorem \ref{Th-3}. }
\end{theorem}

\noindent{\it Proof.} The proof will be based on  Theorem
VIII.25 (a) from \cite{ReedSimon2}: Let {  $\{A_n\}^\infty_{n=1}$ and $A$}
be self-adjoint operators and suppose that $\cD$ is a common core for
all $A_n$ and $A$. If $A_n\phi \rightarrow A\phi$ for each $\phi \in
\cD$, then $A_n\rightarrow A$ in the strong resolvent sense.

{   We start with the case $m^2\neq\frac14$, so
  that we can assume that $f_\Lambda=f_{m,\Tilde{\kappa},\mathrm{D}}$ and 
  $g_\Lambda=g_{m,\Tilde{\kappa},\mathrm{N}}$.
 We can also use the description of maximal operators and
  self-adjoint realizations of Theorem  \ref{Th-3}. }

Consider first the Dirichlet case. Fix vectors $\psi_{\pm 
  m,\mathrm{D}}$, as in \eqref{pqw1}. We will use
the space
\begin{equation}\cD=L_1^{2,\infty}(\rr_+)+\mathbb{C}(\psi_{
  m,\mathrm{D}}+\Tilde\kappa \psi_{
  - m,\mathrm{D}}),\end{equation}
which is a core of
$\Tilde{H}_{m,\Tilde{\kappa},\mathrm{D}}$.
If $\psi_{\min}\in L_1^{2,\infty}(\rr_+)$, then clearly
\begin{equation}
  \Tilde{  L}_{m^2,\mathrm{D}}(\Lambda)\psi_{\min}\, \underset{\Lambda\to\infty}\to\,
  \Tilde{  L}_{m^2,\mathrm{D}}^{\min}\psi_{\min}=
    \Tilde{H}_{m,\Tilde{\kappa},\mathrm{D}}\psi_{\min}.\end{equation}
  Moreover,
  \begin{align}
    K_\mathrm{D}(\Lambda)\psi_{\min}(p)
    =-\theta(\Lambda-p) p\int_\Lambda^\infty q\psi_{\min}(q)\mathrm{d} q.
  \end{align}Now $f_{m,\Tilde{\kappa},\Lambda}$ is uniformly bounded,
\begin{equation}
  \int_\Lambda^\infty q\psi_{\min}(q)\mathrm{d} q\,=\,
  O(\Lambda^{-\infty}),\qquad
\Big(  \int_0^\infty\theta(\Lambda-p)p^2\mathrm{d}p\Big)^\frac12=O(\Lambda^{\frac32}).
   \end{equation}
Therefore,
\begin{equation}
  \frac{f_{m,\Tilde{\kappa},\Lambda}}{\Lambda}
K_\mathrm{D}(\Lambda) \psi_{\min} \, \underset{\Lambda\to\infty}\to\,0.
\end{equation}

  Now consider $\psi:=\psi_{m,\mathrm{D}}+
\Tilde\kappa \psi_{
  - m,\mathrm{D}}$. Remember that $f_\Lambda$ is chosen such that
\eqref{jk1} is $0$. Therefore, for large enough $\Lambda$,
  \begin{align}
    \frac{f_\Lambda}{\Lambda}K_\mathrm{D}(\Lambda)\psi(p)
=&\theta(\Lambda-p)    \frac{f_\Lambda}{\Lambda}p
\Big(\frac{\Lambda^{\frac12-m}}{(\frac12-m)}
+\Tilde\kappa\frac{\Lambda^{\frac12+m}}{(\frac12+m)}\Big)
    \\=&\theta(\Lambda-p)  
         \Big(m^2-\frac14\Big)p\Big(\frac{\Lambda^{-\frac12-m}}{(-\frac12-m)}
         +\Tilde\kappa\frac{\Lambda^{-\frac12+m}}{(-\frac12+m)}\Big).
\end{align}Hence,
\begin{align}&
\Big(   \Tilde{  L}_{m^2,\mathrm{D}}(\Lambda)+\frac{f_{m,\Tilde{\kappa},\Lambda}}{\Lambda}\Big)\psi(p)
  \\
  =&\theta(\Lambda-p)\left(
     p^2\psi(p)+\Big(m^2-\frac14\Big) 
     \Bigg(
    \int_0^\Lambda\mathrm{d}q\big(\theta(p-q)q 
    +\theta(q-p)p\big)\psi(q) 
+\frac{p\Lambda^{-\frac12-m}}{\frac12+m}
    +\frac{\Tilde{\kappa}p\Lambda^{-\frac12+m}}{\frac12-m}
     \Bigg)\right)\\
    =&\theta(\Lambda-p)\left(
     \big(p^2(\psi(p)-p^{-\frac32- m}-\Tilde{\kappa}p^{-\frac32+m}\big)+\Big(m^2-\frac14\Big) 
\int_p^\infty\mathrm{d} q(p-q)\big(\psi(q)-q^{\frac32- m}-\Tilde{\kappa}q^{-\frac32+m}\big)\right).
\end{align}
We can drop $\theta(\Lambda-p)$ from the expression above (remember that $\Lambda$ is  
large enough) and we obtain  
 \begin{align}
\Big(   \Tilde{  L}_{m^2,\mathrm{D}}(\Lambda)+\frac{f_{m,\Tilde{\kappa},\Lambda}}{\Lambda}K_\mathrm{D}(\Lambda)\Big)\psi
\underset{\Lambda\to\infty}\to
       \Tilde{  L}_{\alpha,\mathrm{D}}^{\max}\psi
       =\Tilde{  H}_{m,\Tilde{\kappa},\mathrm{D}}\psi.
\end{align}

The Neumann case is analogous. Fix vectors $\psi_{\pm 
  m,\mathrm{N}}$, as in \eqref{pqw2}. We will use
the space
\begin{equation}\cD=L_0^{2,\infty}(\rr_+)+\mathbb{C}(\psi_{
  m,\mathrm{N}}+\Tilde\kappa \psi_{
  - m,\mathrm{N}}),\end{equation}
which is a core of
$\Tilde{H}_{m,\Tilde{\kappa},\mathrm{N}}$.
If $\psi_{\min}\in L_0^{2,\infty}(\rr_+)$, then clearly
\begin{equation}
  \Tilde{  L}_{m^2,\mathrm{N}}(\Lambda)\psi_{\min}\, \underset{\Lambda\to\infty}\to\,
  \Tilde{  L}_{m^2,\mathrm{N}}^{\min}\psi_{\min}=
    \Tilde{H}_{m,\Tilde{\kappa},\mathrm{N}}\psi_{\min}.\end{equation}
  Moreover,
  \begin{align}
    K_\mathrm{N}(\Lambda)\psi_{\min}(p)
    =-\theta(\Lambda-p) \int_\Lambda^\infty \psi_{\min}(q)\mathrm{d} q.
  \end{align}Now, $g_{m,\Tilde{\kappa},\Lambda}$ is uniformly bounded,
\begin{equation}
  \int_\Lambda^\infty \psi_{\min}(q)\mathrm{d} q\,=\,
  O(\Lambda^{-\infty}),\qquad
\Big(  \int_0^\infty\theta(\Lambda-p)\mathrm{d}p\Big)^\frac12=O(\Lambda^{\frac12}).
   \end{equation}
Therefore,
\begin{equation}
\Lambda g_{m,\Tilde{\kappa},\Lambda}
K_\mathrm{N}(\Lambda) \psi_{\min} \, \underset{\Lambda\to\infty}\to\,0.
\end{equation}

  Now consider $\psi:=\psi_{m,\mathrm{N}}+
\Tilde\kappa \psi_{
  - m,\mathrm{N}}$. Remember that $g_\Lambda$ is chosen such that
\eqref{jk2} is $0$. Therefore,
  \begin{align}
    \Lambda g_\Lambda K_\mathrm{N}(\Lambda)\psi(p)
=&\theta(\Lambda-p)    \Lambda g_\Lambda 
\Big(\frac{\Lambda^{-\frac12-m}}{(-\frac12-m)}
+\Tilde\kappa\frac{\Lambda^{-\frac12+m}}{(-\frac12+m)}\Big)
    \\=&\theta(\Lambda-p)  
         \Big(m^2-\frac14\Big)\Big(\frac{\Lambda^{\frac12-m}}{(\frac12-m)}
    +\Tilde\kappa\frac{\Lambda^{\frac12+m}}{(\frac12+m)}\Big) .
\end{align}
Hence,
\begin{align}&
\Big(   \Tilde{  L}_{m^2,\mathrm{N}}(\Lambda)+\Lambda g_{m,\Tilde{\kappa},\Lambda}\Big)\psi(p)
  \\
  =&\theta(\Lambda-p)\left(
     p^2\psi(p)-\Big(m^2-\frac14\Big) 
     \Bigg(
    \int_0^\Lambda\mathrm{d}q\big(\theta(p-q)p
    +\theta(q-p)q\big)\psi(q)
-\frac{\Lambda^{\frac12-m}}{\frac12-m}
    -\frac{\Tilde{\kappa}\Lambda^{\frac12+m}}{\frac12+m}    \Bigg)\right).
\end{align}
Taking into account \eqref{mm2o}, we obtain
\begin{align}
\Big(   \Tilde{  L}_{m^2,\mathrm{N}}(\Lambda)+\Lambda g_{m,\Tilde{\kappa},\Lambda}K_\mathrm{N}(\Lambda)\Big)
  \psi
\underset{\Lambda\to\infty}\to
       \Tilde{  L}_{\alpha,\mathrm{N}}^{\max}\psi
       =\Tilde{  H}_{m,\Tilde{\kappa},\mathrm{N}}\psi.&& 
\end{align}

{   Consider now the case $\alpha=\frac14$. In this case $\kappa$ takes a different role than in the previous part: taking $\lambda_{\mathrm{D}} = \frac{\pi}{2\kappa}$ and $\lambda_{\mathrm{N}} = \frac{2\kappa}{\pi}$ in formulas \eqref{eq-f-Lambda} and \eqref{eq-g-Lambda} we get
\begin{align}
& \frac{f_{\frac12, \underline{\kappa}, \Lambda}}{\Lambda} = -\frac{1}{\Lambda + \frac{\pi}{2\kappa}},\\
&\Lambda  g_{\frac12, \underline{\kappa}, \Lambda} = \frac{1}{\frac{\pi}{2\kappa} + \Lambda^{-1}}.
\end{align}
Take $\cD\big( \Tilde{H}_{\pm\frac12,\underline\kappa, \mathrm{D}} \big) \ni \psi_{\mathrm{D}} = \psi_{\mathrm{D}}^{\mathrm{min}} + \Tilde{c}\psi_{\mathrm{D}, \underline{\kappa}}$ and  $\cD\big( \Tilde{H}_{\pm\frac12,\underline\kappa, \mathrm{N}} \big) \ni  \psi_{\mathrm{N}} = \psi_{\mathrm{N}}^{\mathrm{min}} + \Tilde{c}\psi_{\mathrm{N}, \underline{\kappa}}$, where $\psi_{\mathrm{D/N}}^{\mathrm{min}}$ is in the minimal domain and $\Tilde{c} \in \mathbb{C}$. Using definitions \eqref{pqw1ou}-\eqref{pqw2ou} one checks that for large enough $\Lambda$
\begin{align}
	&K_{\mathrm{D}}(\Lambda)  \psi_{\mathrm{D}}(p) = \theta(\Lambda - p) \Tilde{c} \Big( \Lambda + \frac{\pi}{2\kappa} \Big) p + O(\Lambda^{-\infty}),\\
	&K_{\mathrm{N}}(\Lambda)  \psi_{\mathrm{N}}(p) = -\theta(\Lambda - p) \Tilde{c} \Big( \frac{\pi}{2\kappa} + \Lambda^{-1} \Big) + O(\Lambda^{-\infty}).
\end{align}
Hence, from formulas \eqref{mm1oo}-\eqref{mm2oo} we obtain
\begin{align}
&\Big(   \Tilde{  L}_{\frac14,\mathrm{D}}(\Lambda)+ \frac{f_{\frac12, \underline{\kappa}, \Lambda}}{\Lambda} K_\mathrm{D}(\Lambda)\Big)
  \psi
\underset{\Lambda\to\infty}\to
       \Tilde{  L}_{\frac14,\mathrm{D}}^{\max}\psi
       =\Tilde{  H}_{\frac14,\underline{\kappa},\mathrm{D}}\psi, \\
&\Big(   \Tilde{  L}_{\frac14,\mathrm{N}}(\Lambda)+\Lambda g_{\frac12, \underline{\kappa} } K_\mathrm{N}(\Lambda)\Big)
  \psi
\underset{\Lambda\to\infty}\to
       \Tilde{  L}_{\frac14,\mathrm{N}}^{\max}\psi
       =\Tilde{  H}_{\frac14,\underline{\kappa},\mathrm{N}}\psi.
\end{align}
The case $m = -\frac12$ is treated in the analogous way, by taking $f_{-\frac12, \underline{\kappa} } = f_{\frac12, \underline{\kappa}^{-1} }$ and $g_{-\frac12, \underline{\kappa} } = g_{\frac12, \underline{\kappa}^{-1} }$.
  $\Box$}

Let us observe, that the
norm of the counter-terms converges to infinity both in the Dirichlet
and Neumann cases. In particular, we cannot neglect the counterterms
even in the Dirichlet case with $|\mathrm{Re}(m)| < \frac12$ (when one
can omit the counter-terms in Eq. \eqref{mm1o}).

Note also  that the above analysis 
of   $\Tilde{H}_{m,\Tilde{\kappa},\mathrm{D}}$ is essentially an expanded version
of  the Wilsonian approach of \cite{Glazek}, which we described in
Section \ref{sec2},
translated from 3 dimensions to  1 dimension.

\section{Conclusion}
%\init
Treatment of the Schr\"odinger equation with potential $1/r^2$ requires an appropriate definition of the  domain. This can be achieved directly, in the position representation, as it was presented in the Section 2. It can be also equivalently done in the momentum representation,  essentially following the Wilsonian renormalization scheme. Despite having been devised as an approximate method, this scheme when rigorously implemented yields a construction of self-adjoint realizations of the Schr\"odinger operator 
with potential $1/r^2$.

\renewcommand{\abstractname}{Acknowledgements}

\begin{abstract}
  O.G. would like to thank Prof. Stanisław Głazek for suggesting this research topic and his comments on this work. He  also thanks Maciej Łebek and Ignacy Nał\c{e}cz for discussions about the Wilsonian approach.

  J.D. acknowledges useful discussions about renormalization with Prof. Stanisław Głazek. The work of J.D. was supported by  National Science Center (Poland) under the grant UMO-2019/35/B/ST1/01651.
\end{abstract}

\section{Author Declarations}
\subsection{Data Availability Statement}
 The data that supports the findings of this study are available within the article.
\subsection{Conflict of interest}
The authors have no conflicts to disclose.

\appendix

  \section{Fourier analysis on a halfline} \label{appx}
%\init

Fourier analysis on the line is well-known. Somewhat less 
  known is Fourier analysis on the half-line, where the role of the 
  Fourier transformation is played by two transformations: the 
cosine and sine transformation, which are the main subject of this
appendix.

  \subsection{Cosine and sine transformation}

Let us start with recalling some aspects of 
Fourier analysis on the line.
 Let
\begin{align}
H^{2,\infty}(\rr)&:=\Big\{ \phi\in C^\infty(\rr)\ |\ \int_{-\infty}^\infty 
  |\phi^{(n)}(x)|^2\mathrm{d} x<\infty,\quad
                     n=0,1,\dots\Big\},\label{frech1}\\
  L^{2,\infty}(\rr)&:=\Big\{\psi\in L^2(\mathbb{R})\ |
      \int_{-\infty}^\infty|\psi(p)|^2|p|^n\mathrm{d}p<\infty,\quad
                     n=0,1,\dots\Big\}
                     \label{frech2}
\end{align}
be the Sobolev space and the weighted space of the infinite order--two
examples of Frechet
spaces.
The Fourier transformation swaps these spaces:
  \begin{align}
                      \cF L^{2,\infty}(\rr)&=H^{2,\infty}(\rr).
                                             \end{align}

 \noindent Functions on the line can be decomposed
into even and odd functions
  \begin{equation}L_\pm^2(\mathbb{R}):=\{\psi\in L^2(\mathbb{R})\ |\
    \psi(-x)=\pm\psi(x)\},\quad L^2(\rr)=L_+^2(\rr)\oplus L_-^2(\rr).\end{equation}
 Even and odd functions are preserved by  the Fourier transformation:
  \begin{align}\label{posd}
    \cF L_\pm^2(\rr)&= L_\pm^2(\rr)%,\\
                      %\cF L^{2,\infty}(\rr)&=H^{2,\infty}(\rr).
                                             \end{align}

\noindent
Every function on the halfline can be extended to an even or odd function: 
  \begin{equation}  J_\pm\psi(p):=
    \left\{
	\begin{array}{ll}
	\psi(p) \quad & p \geq 0 \\
	\pm \psi(-p)  \quad & p < 0.
	\end{array}
	\right.
\end{equation}
$J_\pm$ maps $L^2(\rr_+)$ onto $L_\pm^2(\rr)$. The restriction to the
positive halfline is the left inverse of $J_\pm$:
\begin{equation}\label{restri}
(J_\pm\psi)\Big|_{\rr_+}=\psi,\quad \psi\in L^2(\rr_+).
\end{equation}
The sine and cosine transformations can be defined
as the composition of the Fourier transformation with the above
extension and the restriction, more precisely,
\begin{equation}
  \cF_\mathrm{D}\psi=\mathrm{i}( \cF J_-\psi)\Big|_{\rr_+}
  ,\quad    \cF_\mathrm{N}\psi= (\cF J_+\psi)\Big|_{\rr_+}.\end{equation}

Introduce also the  Frechet spaces analogous to \eqref{frech1} and
\eqref{frech2} corresponding to the halfline:
\begin{align}
H^{2,\infty}(\rr_+)&:= \Big\{ \phi\in C^\infty(\rr_+)\ |\ \int_0^\infty
  |\phi^{(n)}(x)|^2\mathrm{d} x<\infty,\quad
                     n=0,1,\dots\Big\},\\
                 L^{2,\infty}(\rr_+)&:=\Big\{\psi\in L^2(\mathbb{R}_+)\ | \int_0^\infty|\psi(p)|^2p^n \d p<\infty,\quad n=0,1,\dots\Big\}.
\end{align}
We will also need the following  closed subspaces of $ H^{2,\infty}(\rr_+)$:
\begin{align}
 H_+^{2,\infty}(\rr_+)&:= \{\phi\in H^{2,\infty}(\rr_+)   \ |\ \phi^{(n)}=0,\quad n=1,3,5,\dots\}
                        ,\label{closed1}\\
   H_-^{2,\infty}(\rr_+)&:= \{\phi\in H^{2,\infty}(\rr_+)   \ |\ \phi^{(n)}=0,\quad n=0,2,4,\dots\}
                          .\label{closed1-}
\end{align}

The following proposition is straightforward:

  \begin{proposition}
    \begin{align}
  J_+ H_+^{2,\infty}(\rr_+)&=
                             H^{2,\infty}(\rr)\cap L_+^2(\rr),&
                                                                  J_- H_-^{2,\infty}(\rr_+)&=
  H^{2,\infty}(\rr)\cap L_-^2(\rr),\\ 
      \cF_\mathrm{N} L^{2,\infty}(\rr_+)&= H_+^{2,\infty}(\rr_+),&            \cF_\mathrm{D} L^{2,\infty}(\rr_+)&=
                                                                                                                  H_-^{2,\infty}(\rr_+).\label{formu2}
                                                               \end{align}
                                                               For $\psi\in L^{2,\infty}(\rr_+)$ we have
                                                               \begin{align}
                                                                                  \partial_x^{2n}\cF_\mathrm{D}\psi&=(-1)^n 
                                                                                                                     \cF_\mathrm{D}p^{2n}\psi,&                                                                                  \partial_x^{2n}\cF_\mathrm{N}\psi&=(-1)^n 
                                                                                                    \cF_\mathrm{N}p^{2n}\psi,\label{formu3}\\
                                                                 \partial_x^{2n+1}\cF_\mathrm{D}\psi&
                                                                                                      =(-1)^{n+1} 
                                                                                                                     \cF_\mathrm{N}p^{2n+1}\psi,&                                                                                  \partial_x^{2n+1}\cF_\mathrm{N}\psi&=(-1)^n 
                                                                                                                                                                                                                                                                    \cF_\mathrm{D}p^{2n+1}\psi.
         \label{formu4}                                                      \end{align}
                                                               
                                                             \end{proposition}

\subsection{Homogeneous functions on the halfline}

Let $f\in L_\mathrm{loc}^1(\rr_+)$. We say that $f$ possesses an
oscillatory integral if for any $\phi\in
C_\mathrm{c}^\infty([0,\infty[)$ such that $\phi=1$ near $0$
\begin{equation}\lim_{\Lambda\to\infty}\int_0^\infty
  f(p)\phi(p/\Lambda)\mathrm{d}p\label{osci}\end{equation}
exists and does not depend on the choice of $\phi$. The value 
\eqref{osci} is then called the {\em oscillatory integral of $f$}.

Note that the integrals that appear in the definitions of the cosine
and sine transforms \eqref{fd}, \eqref{fn} for functions, say, from
$L^2(\rr_+)$ can  always be understood as oscillatory
(but not always in the  usual Lebesgue sense).

In   the following
formulas, valid for $x>0$, one needs to use oscillatory
integrals for $\lambda>-1$:
  \begin{align}
    \sqrt{\frac{2}{\pi}}\int_0^\infty\sin(px) x^\lambda\mathrm{d}x&=
-\sqrt{\frac\pi2}\frac{p^{-\lambda-1}}{\sin(\frac\pi2 \lambda)\Gamma(-\lambda)},\quad\lambda>-2   \label{eq-sin-tr} ,\\
\sqrt{\frac{2}{\pi}}\int_0^\infty\cos(px) x^\lambda\mathrm{d}x&=
\sqrt{\frac\pi2}\frac{p^{-\lambda-1}}{\cos(\frac\pi2 \lambda)\Gamma(-\lambda)},\quad\lambda>-1 . \label{eq-cos-tr}
\end{align}

\begin{lemma} Suppose that $\psi\in L_\mathrm{loc}^1(\rr_+)$ such
  that for large $p$ we have $\psi(p)=p^\lambda$. Set
  \begin{align}
    \psi_1(p)&:=\lim_{\Lambda\to\infty}\Big(-\int_p^\Lambda 
               \psi(q)\mathrm{d}q+\frac{\Lambda^{\lambda+1}}{\lambda+1}\Big),\\
        \psi_2(p)&:=\lim_{\Lambda\to\infty}\Big(\int_p^\Lambda 
                   (p-q)\psi(q)\mathrm{d}q-\frac{p\Lambda^{\lambda+1}}{\lambda+1}
                   +\frac{\Lambda^{\lambda+2}}{\lambda+2}\Big).
    \end{align}
(The above limits exist, because the functions after the limit sign are
constant for large $\Lambda$).
Clearly,
\begin{align}
  &\psi_2'(p)=\psi_1(p),\qquad \psi_1'(p)=\psi(p),\\
\text{ for large $p$,}\quad &
                                \psi_1(p)=\frac{p^{\lambda+1}}{\lambda+1},\quad
                                                                \psi_2(p)=\frac{p^{\lambda+2}}{(\lambda+1)(\lambda+2)}. \label{eq-psi12-large-p}
\end{align}
Moreover, the following holds:
\begin{enumerate}
\item
  Suppose that
  \begin{equation}\label{bdary1}
    \lim_{\Lambda\to\infty}\Big(-\int_0^\Lambda 
   q
   \psi(q)\mathrm{d}q+\frac{\Lambda^{\lambda+2}}{\lambda+2}\Big)=0.
 \end{equation}
 Then
 \begin{align}
   \int_0^\infty\sin(px)\psi(p)\mathrm{d}p=
-   x   \int_0^\infty\cos(px)\psi_1(p)\mathrm{d}p=
 -  x^2   \int_0^\infty\sin(px)\psi_2(p)\mathrm{d}p.
 \end{align}
\item
  Suppose that
  \begin{equation}\label{bdary2}
    \lim_{\Lambda\to\infty}\Big(\int_0^\Lambda 
   \psi(q)\mathrm{d}q-\frac{\Lambda^{\lambda+1}}{\lambda+1}\Big)=0.
 \end{equation}
 Then
 \begin{align}
   \int_0^\infty\cos(px)\psi(p)\mathrm{d}p=
   x   \int_0^\infty\sin(px)\psi_1(p)\mathrm{d}p=
 -  x^2   \int_0^\infty\cos(px)\psi_2(p)\mathrm{d}p.
 \end{align}
\end{enumerate}
(The above integrals are not always defined as Lebesgue 
integrals---they are always defined as oscillatory integrals).
\label{ppen}
\end{lemma}

\noindent{\it Proof.}  1.
Let $\phi$ be as in the definition of the oscillatory integral.
We integrate by parts:
 \begin{align}
  & \int_0^\infty\phi(p/\Lambda)\sin(px)\psi(p)\mathrm{d}p\\=&
                                                -   x   \int_0^\infty\phi(p/\Lambda)\cos(px)\psi_1(p)\mathrm{d}p 
                                                 -\Lambda^{-1}\int_0^\infty\phi'(p/\Lambda)\sin(px)\psi_1(p)\mathrm{d}p 
   \\
                                                =&\label{eq-phi-prime}
                                                   -  x^2
                                                   \int_0^\infty\phi(p/\Lambda)\sin(px)\psi_2(p)\mathrm{d}p
                                                                                                 -\Lambda^{-1}\int_0^\infty\phi'(p/\Lambda)\sin(px)\psi_1(p)\mathrm{d}p \\ &
                                                   -x\psi_2(0)
                                                   +x\Lambda^{-1}\int_0^\infty\phi'(p/\Lambda)\cos(px)\psi_2(p)\mathrm{d} p.
 \end{align}
 Now $\psi_2(0)=0$ because of \eqref{bdary1}.
 { Terms involving $\phi'$
 are $O(\Lambda^{-\infty})$, what can be checked by integration by parts}. %For example, consider the second term in
% \eqref{eq-phi-prime}. Substituting $y = p/\Lambda$,
%using $\supp(\phi') \subset \: ]0,\infty[$ and \eqref{eq-psi12-large-p}, we obtain that for large $\Lambda$ this integral equals
%\begin{equation}
%	\frac{\Lambda^{\lambda+1}}{\lambda + 1}\int_0^\infty  \phi'(y) y^{\lambda+1} \sin(xy\Lambda)\d y.\label{integr}
%\end{equation}
%Let us set
%\[
%\varphi(y) = \left\{
%					\begin{array}{ll}
%						\phi'(y) y^{\lambda+1} \quad &y \geq 0,\\
%						0 \quad & y < 0,
%					\end{array}
%\right.
%\]
%so that the integral in \eqref{integr} is the imaginary part of the
%  Fourier transform of  $\varphi \in C^\infty_\mathrm{c}(\mathbb{R})$ evaluated
%  at $x\Lambda$. This Fourier transform is a Schwartz function.
% Therefore, 
%  for $x\neq 0$ \eqref{integr} is $O(\Lambda^{-\infty})$.
  
2. The proof is similar:
 \begin{align}
  & \int_0^\infty\phi(p/\Lambda)\cos(px)\psi(p)\mathrm{d}p\\=&
                                                 x   \int_0^\infty\phi(p/\Lambda)\sin(px)\psi_1(p)\mathrm{d}p 
                 -\psi_1(0)                                +\Lambda^{-1}\int_0^\infty\phi'(p/\Lambda)\cos(px)\psi_1(p)\mathrm{d}p 
   \\
                                                =&
                                                   -  x^2
                                                   \int_0^\infty\phi(p/\Lambda)\cos(px)\psi_2(p)\mathrm{d}p
                   -\psi_1(0)                                                                              +\Lambda^{-1}\int_0^\infty\phi'(p/\Lambda)\cos(px)\psi_1(p)\mathrm{d}p \\&
                                                                      +x\Lambda^{-1}\int_0^\infty\phi'(p/\Lambda)\sin(px)\psi_2(p)\mathrm{d} p.
 \end{align}
 We use \eqref{bdary2} to get $\psi_1(0)=0$.
Again, the  terms  involving $\phi'$
 are $O(\Lambda^{-\infty})$.
 \qed

\end{document}